# Fast Dynamics in a Model Metallic Glass-forming Material

Hao Zhang[1†], Xinyi Wang[1], Hai-Bin Yu[2], Jack F. Douglas[3†]

[1] Department of Chemical and Materials Engineering, University of Alberta, Edmonton, Alberta, Canada, T6G 1H9

[2] Wuhan National High Magnetic Field Center, Huazhong University of Science and Technology, Wuhan, Hubei, China, 430074

[3] Material Measurement Laboratory, Material Science and Engineering Division, National Institute of Standards and Technology, Maryland, USA, 20899

[†]Corresponding authors: hao.zhang@ualberta.ca; jack.douglas@nist.gov

**Abstract**

We investigate the fast $\beta$- and Johari-Goldstein (JG) $\beta$-relaxation processes, along with the elastic scattering response of glass-forming (GF) liquids and the Boson peak, in a simulated Al-Sm GF material exhibiting a fragile-strong (FS) transition. These dynamical processes are universal in 'ordinary' GF fluids and collectively describe their 'fast dynamics', and we find these relaxation processes also arise in a GF liquid exhibiting a fragile-strong transition. String-like particle motion, having both an irreversible and reversible nature ('stringlets') component, occurs in the fast-dynamics regime, corresponding to a ps timescale. String-like collective motion associated with localized *unstable* modes facilitate irreversible and intermittent particle 'jumping' events at long times associated with the JG $\beta$-relaxation process, while stringlets associated with localized stable modes and corresponding perfectly reversible atomic motion give rise to the Boson peak. To further clarify the origin of the Boson peak, we calculate the density of states for both the stringlet particles and the 'normal' particles and find that the stringlet particles give rise to a Boson peak while the normal atoms do not. The growth of stringlets upon heating ultimately also leads to the 'softening' of these excitations, and the Boson peak frequency and shear modulus drop in concert with this softening. The growth of string-like collective motion upon *heating* in the fast-dynamics regime is further shown to be responsible for the growth in the intensity of the fast relaxation process. Relaxation in cooled liquids clearly involves a hierarchy of relaxation processes acting on rather different time and spatial scales.

## I.    Introduction

While the dramatic slowing down the 'primary' or '$\alpha$-structural relaxation time' $\tau_\alpha$ of glass-forming (GF) materials has received the most scientific attention because of its obvious relevance to applications, relaxation in GF liquids involves a *hierarchy* of relaxation processes that are somehow linked to each other, but their interrelations are currently obscure given the absence of a general organizing theory of glass-formation. Much information about glasses has emerged from an extensive body of experimentation and numerous computational studies, however. In particular, many measurements and simulations have revealed the general occurrence of a 'fast' relaxation process in all molecular GF liquids for which the relaxation time $\tau_f$ is typically on the order of a ps and exhibits weak or no apparent $T$ dependence. [1] This relaxation process is often termed the 'fast $\beta$-relaxation' process, which must be differentiated from the Johari-Goldstein (JG) $\beta$-relaxation or 'slow $\beta$-relaxation' process for which the relaxation time $\tau_{JG}$ is typically in the range [1] of $10^{-9}$ s to $10^{-7}$ s and exhibits an Arrhenius $T$ dependence in the glass state. Since the $\alpha$-relaxation time $\tau_\alpha$ normally becomes longer than minutes below the glass transition temperature $T_g$, this channel of relaxation largely 'shuts down' in the glass state, and faster relaxation processes by default then become of preeminent importance for understanding the performance properties of glass materials for many applications. The lack of understanding of the molecular origin of $\beta$-relaxation processes is then a problem of both theoretical and technological concern. In a previous paper, we systematically investigated $\tau_\alpha$ and the diffusion coefficient $D$ in a model Al-Sm metallic GF liquid exhibiting a relatively sharp glass transition of the fragile-strong (FS) type. [2] The occurrence of this relatively abrupt glass-formation facilitates molecular dynamics simulation of the transition because a much more limited $T$ range is required. The material we study is also special because it also exhibits a pronounced JG $\beta$-relaxation that is well separated from the $\alpha$- relaxation process. Moreover, this system has been well studied, allowing for the validation of the many-body atomic potential used in our simulations from the standpoint of static scattering and other thermodynamic measurements. [3-6] Importantly for the present work, there is excellent experimental data for the JG $\beta$-relaxation of this model metallic class material and recent molecular dynamics simulations have been able to reproduce these observations to a remarkable degree. [3] This is a significant computational accomplishment because simulations of large systems for timescales as long as microseconds, or longer, are required to compare with the JG $\beta$-relaxation measurements.

Based on this prior work, we are positioned to quantify the fast relaxation processes in this model material and to explore possible relations of these relaxation processes to dynamical heterogeneities previously identified in simulations of the $\alpha$-relaxation process. Our main, and admittedly optimistic, intent is to obtain a more holistic description of all the relaxation processes in this model GF liquid that might form the basis for the development of a more unified understanding GF materials broadly. We also address the origin of another fast dynamics process, the boson peak, typically occurring on a timescale in order of magnitude shorter than the fast relaxation time $\tau_f$, and find that this scattering feature corresponds on an atomic scale to particles undergoing string-like collective motion that is completely reversible, as opposed to the fast relaxation process where the particle motions are likewise dominated by particles exhibiting large scale string-like motion that involves irreversible motion. The spatially limited and reversible nature of the normal mode motions of the atoms giving rise to the Boson peak strongly suggested to us that these localized vibrational modes are distinct from phonons, and we correspondingly verified the distinct character of these normal modes by calculating the density of states associated with these atoms from their velocity autocorrelation function. The atoms involved in these string-like reversible collective motions exhibit a strong Boson peak feature in their density of states, which is not consistent with these modes being of a phonon nature. Moreover, we did not find this 'anomaly' in the density of states of representatively chosen atoms that were not undergoing reversible collective motion. We take this as 'smoking gun' evidence of the normal mode motions responsible Boson peak, at least in the present material, and we discuss the implications of this type localized stable mode of the light-vibrational coupling and other aspects of the Boson peak phenomenon relevant to the observation of the Boson peak in neutron and light scattering measurements broadly. It is implicit in this discussion that the fast dynamics string-like collective motion that we observe in our metallic glass is universal to *all* glass-forming liquids. This hypothesis remains to be proved by studying other glass-forming liquids using similar methods.

This type of collective excitation is found to have a direct relation to understanding the origin of the fast $\beta$- and the JG $\beta$-relaxation processes operating on longer timescales. In particular, the growth of the amplitude of the local atomic motion associated with these localized modes, and the increasing amplitude of the displacements associated with these modes upon heating, are found to greatly contribute to the growth of the intensity of the fast $\beta$-relaxation process upon heating, and these large amplitude motions also facilitate the intermittent jumping motions underlying the JG $\beta$-relaxation process on much longer times where the stringlet excitations (particles undergoing cooperative exchange motion in the form of strings on a ps timescale) become overdamped and disintegrate. The full range of relaxation processes is apparently 'fed' by these collective excitations, which apparently also have significant relevance for understanding the low temperatures  properties such as the specific heat and thermal conductivity of GF materials and for understanding the mysterious two-level systems phenomenologically invoked to rationalize the universal low temperature ($T$) properties of glassy materials. We establish a direct relation between the stringlets and Boson peak by calculating the density of states of the stringlet atoms and unexcited 'normal' atoms, as well as the density of states of the entire metallic glass material. The particle in the stringlet mode give rise to a Boson peak while the normal particles do not. The entire material exhibits a Boson peak, so this feature arises from the stringlets. Finally, we study the softening of the Boson peak and the corresponding reduction of the bulk and shear modulus of the material and interrelate these phenomena.

After defining the model and computational methods briefly in Sect. II, we summarize our observations on the fast $\beta$-relaxation time $\tau_f$ and fundamentally important fast dynamics property, the mean-square displacement $<u^2>$ defined on ps 'caging time' $\tau_{f\beta}$ or 'Debye-Waller factor'. We show that there is significant dynamic heterogeneity in the fast dynamics regime involving highly coordinated string-like collective atomic motion of an inertial origin ('stringlets'), whose scale grows upon heating rather than cooling. These collective excitations make a very large contribution to $<u^2>$ of the entire system, even though relatively few particles are activated into this highly mobile particle state. The rapid increase of $<u^2>$ with increasing temperature and the corresponding decrease of the amplitude of the fast $\beta$-relaxation process $A_\beta$ in comparison with the $\alpha$-relaxation process $1 - A_\beta$ can also be traced to this fast-collective motion. We also find that the relaxational contribution to the fast dynamics involves a complementary component involving again large amplitude string-like displacements that are *completely reversible*, corresponding to stable normal modes, and we confirm from the calculation of the Boson peak of atoms involved in these motion that gives rise to the Boson peak, while other atoms do not give rise to any such contribution to the density of states. We suggest that this phenomenon arises from a bifurcation of the localized stable modes found for a $T = 0$ inherent structure analysis into stable and unstable localized modes at a finite $T$, just as the stable and unstable modes separate from their stable mode

"parents" at finite temperature. The degree of separation of these modes depends on the stability of these modes, and thus $T$, all the normal modes becoming unstable upon approaching the high temperature gas state and all the modes being stable in the opposite extreme of $T = 0$. The particles undergoing this collective motion exhibit physical characteristics radically different from the surrounding particles, so it is easy to identify these particles. In particular, the local potential energy of these particles undergoes discrete jumps between three distinct potential energy states that seems the potential energy jumps are accompanied by large jumps in particle displacement whose statistics resembles earthquake data in that the magnitude of $<u^2>$ fluctuates with a power-law distribution. We examine the local structural changes (local coordination changes and shape changes in the Voronoi cells defined by each atom) in the vicinity of these jumps to better understand this striking phenomenon, which seems to be facilitated by the stringlet motion that dominates the fast relaxation process. An analysis of the potential energy fluctuations of the mobile particles reveals that these fluctuations exhibit colored noise where the noise exponent peaks in magnitude as the same temperature as the specific heat, isothermal compressibility the $\chi_4$ function measuring the intensity of mobility fluctuations, a pattern of behavior exactly parallel to previous observations on supercooled water. [7,8] This analysis of the fast dynamics culminates with the observation that the jumps occur intermittently, but with an average rate consistent with Johari-Goldstein relaxation time, $\tau_{JG}$. We also find that the jump rate time $\tau_{jump}$ also seems to coincide with the lifetime of mobile particle clusters $\tau_M$ identified in previous studies.[3,9,10] This observation is complementary to previous observations that the immobile particle lifetime determines the structural relaxation time $\tau_\alpha$. We also find that $D / T$ scales inversely to both $\tau_{JG}$ and $\tau_{jump}$ over the low $T$ range that we simulate to within numerical uncertainty, consistent with the arguments and observations that have linked JG $\beta$-relaxation process to an inter-basin jump process governing the rate of molecular diffusion in small molecule GF liquids. [11] This scaling also explains the decoupling relation we observe relating $D / T$ to both $\tau_{JG}$ and $\tau_{jump}$. Finally, given that the Boson peak is related to the high elastic response of the material to incident radiation, we show that the Boson peak frequency $\omega_B$ can be related to $<u^2>$, which provides a measure of the high frequency elastic material response.

## II.    Model and Simulation Methods

We employed classical molecular dynamics (MD) simulations to study the fast $\beta$- and JG $\beta$-relaxation processes as well as primary $\alpha$-structural relaxation process of a model metallic GF alloy $Al_{90}Sm_{10}$ using a many-body semi-empirical potential. [12] The potential itself takes the form of Finnis-Sinclair [13], and physical properties of pure aluminum such as cohesive energy, elastic modulus, vacancy formation energy, melting point and crystal structure of a series of Al-rich compounds have been employed for the semi-empirical potential parameters fitting procedure. This semi-empirical potential has proven to be able to reproduce some important thermodynamic and kinetic properties in the previous studies, showing an excellent reproduction of pure Al properties and liquid structure of $Al_{90}Sm_{10}$ at $T = 1273$ K [12,14], predicting accurate short-range order (SRO) structure in the supercooled liquid/glass state[6], and providing reasonable estimation of interfacial free energy for different crystallographic planes in solid-liquid interface [5] In the current study, a liquid sample, containing 28785 Al atoms and 3215 Sm atoms, were initially held at 2000 K for 2.5 ns to reach equilibrium, and then it was continuously cooled down with a constant cooling rate of 0.1 K/ns to 200 K. The periodic boundary conditions were applied in all directions with the isobaric-isothermal ensemble (NPT). The pressure was kept at zero for all simulations and  the simulation box size were controlled by the Parrinello-Rahman method[15] and constant temperature $T$ was maintained by the Nose-Hoover method. [16,17] The MD simulations utilize Large-scale Atomic/Molecular Massively Parallel Simulator (LAMMPS) [18], which was developed at the Sandia National Laboratories. In addition, we also employed isothermal heating for an extended period of time at $T$ = 900 K, 850 K, 800 K, 750 K, 7000 K, 650 K, 600 K, 550 K, 500 K and 450 K to allow us to obtain reasonable equilibrium configurations and to perform structural relaxation analysis later. The simulation time for isothermal heating depends on $T$, and it is typically on the order of tens of ns, but this time is longer than 0.6 μs for low $T$ simulations.

## III.    Results and Discussion

### A.  Fast $\beta$-Relaxation Process

We begin our discussion of the 'fast' relaxation process, which is prevalent in many inelastic neutron and light scattering measurements. In particular, this relaxation process may be identified in any such measurements that allows the determination of the initial decay in self-intermediate scattering function $F_s(q,t)$, whose relaxation time $\tau_f$ in the liquid state is typically on the order of a ps, regardless of temperature for molecular fluids. [1,19] In our Al-Sm metallic glass forming liquid, this time scale is about an order of magnitude smaller, taking a value on the order 0.1 ps, a value consistent with measurements such as inelastic neutron scattering on liquid Na. [20] The measurements are often performed Fourier-Laplace transform of $F_s(q,t)$, but this is just technical detail from the standpoint of the current paper. We summarized the determination of $F_s(q,t)$ and $\tau_f$ in our previous paper, focusing on relaxation in this GF liquid and here we focus on the amplitude of the fast relaxation process corresponding to the intensity of the scattering associated with this process. It is rather universally observed that the intensity of the fast beta relaxation grows linearly, or at least roughly linearly, with $T$ and its intensity tends to extrapolate to 0 around the Vogel-Fulcher-Tammann temperature $T_o$, [21-23] at which a fit to $\tau_\alpha$ relaxation time data above $T_g$ tends to extrapolate towards at lower $T$. While the current GF liquid exhibits a fragile-strong transition [24-27], the VFT equation phenomenology reasonably describes the growth of $\tau_\alpha$ at low $T$ and both $T_g$ and $T_o$ can be formally defined in a conventional fashion to ordinary GF liquids (See SI). Notably, the VFT equation, $\tau_\alpha = \tau_0 \exp\left[\frac{DT_0}{T-T_0}\right]$, [21-23] does not provide a good description of the relaxation time in the high T regime of glass-formation defined by a crossover temperature $T_c$ (We provide a complete characterization of all the characteristic temperatures for our metallic GF material in the SI of our companion paper [10] focusing on $\tau_\alpha$ and atomic diffusion in our metallic GF material.). This phenomenology has led to the suggestion that the $\alpha$-relaxation might relate to 'free volume', which has been presumed in older models of glass-formation to vanish at $T_o$. [1,19] As we shall discuss below, the fast relaxation has also been correlated with $T_o$, suggesting some sort of link between fast relaxation and $\alpha$-relaxation.

We may gain some insight into the widely observed growth of scattering intensity of the fast $\beta$-relaxation process above $T_o$ from a previous study of the fast $\beta$-relaxation by Betancourt et al. [28] In this work, it was found that the amplitude of the fast $\beta$-relaxation process $A_\beta$, defined to equal 1 minus the amplitude of the $\alpha$-relaxation contribution to the intermediate scattering function, the so-called 'non-ergodicity parameter' $h_{inc}$ can be related to the Debye-Waller factor, the mean square displacement $<u^2>$ of the particles at a caging time on the order of a time on the order of a ps. The caging time is somewhat longer than the fast beta relaxation time $\tau_f$ (often by a factor of 10 or so) describing initial the decay of the $F_s(q,t)$ where the caging time delineates to a transient plateau governing the initiation of the $\alpha$-relaxation process.[29,30] The $\alpha$ and fast $\beta$-relaxation processes then merge at the caging time, which only exists below an onset temperature for caging, $T_A$ (See our companion paper for illustration of the fast $\beta$-relaxation process of our Al-Sm metallic glass [10] and Betancourt et al. [28] for a general discussion of fast $\beta$-relaxation and the determination of the 'caging time', $\tau_{f\beta}$. Historically, $\tau_{f\beta}$, has been identified[28,31-33] in computational studies as defining the fast $\beta$-relaxation process, even though it would seem that $\tau_f$ has a better claim on this term (Coniglio and coworkers [34] have suggested that might be better termed the Debye-Waller time since the Debye-Walker factor $<u^2>$ is defined by the mean square molecular displacement at this same time. We stick to the standard terminology in the present paper to avoid potential confusion.) We include $\tau_{f\beta}$ in our tabulation of characteristic times below.

We next discuss an interesting connection between the fast dynamics and $\alpha$-relaxation. Betancourt et al. found that the non-ergodicity parameter $h_{inc}$ can be described by the approximation, $h_{inc} \approx \exp(- q_o <u^2> / 6)$, where $q_o$ is the scattering wavevector corresponding to the interparticle distance. This is a very good approximation because the Gaussian is a good approximation in this short time regime, at least in this $q$ range. [28] The amplitude $A_\beta$ of the fast beta relaxation (proportional to the scattering intensity of this relaxation process) at low $T$ where $<u^2>$ can be reasonably be approximated as, $A_\beta \equiv 1 - h_{inc} \sim <u^2>$, proving a direct link between the amplitude of the fast relaxation process and the mean amplitude of atomic motion on a ps timescale. This brings us then to a link to $T_o$ since we have repeatedly observed that $<u^2>$ scales as $<u^2> \sim a_o [(T- T_o)/ T_o]^{\delta}$ for $T > T_g$ where we can perform our simulations at equilibrium with $a_o$ a material dependent constant, and the exponent $\delta$ is often near unity, but is sometimes larger depending on the material. [28,35,36] It has been repeatedly observed that the $T$ dependence of $<u^2>$ is stronger in more fragile GF liquids [35,37], corresponding to a larger $a_o$, and/or $\delta$ and, correspondingly $A_\beta$ varies more strongly with $T$ in fragile GF liquids. The roughly linear increase in the amplitude of the fast $\beta$-relaxation process with $T$ at low $T$ near $T_g$ and a tendency of this scattering feature to become essentially extinguished upon extrapolation to $T_o$ is then evidently linked to the sharp increase in the $T$ dependence of $<u^2>$ around $T_g$. An understanding molecular motions underlying this sharp variation of $<u^2>$ with $T$ is then key to understanding the nature of fast relaxation in GF liquids.

Based on this discussion, we may readily understand recent correlations of the amplitude of the $\alpha$-relaxation process, $h_{inc}$, based on X-ray scattering data, with the fragility of glass-formation. We further note that in the glass state, $<u^2>$ can be safely approximated as varying nearly linearly in $T$ so that we have, $h_{inc} = \exp(- q_o <u^2> /6) \approx h^o_{inc} / [1 + \alpha (T / T_o)]$, by expanding the exponential term to leading order and moving this linear term to the denominator, consistent with the leading order expansion in $<u^2>$. In this approximate expression, the constant $\alpha$ absorbs the constant $q_o$, the numerical factor 6, and the parameter is dominated by the constant $a_o$ describing the $T$ dependence of $<u^2>$. Scopigno and coworkers [38] have noted that this phenomenological approximation holds rather well (with $T_o$ replaced by $T_g$) for dynamic X-ray scattering measurements made on a variety of materials where the parameter $\alpha$ increases progressively with the material steepness index $m$, a measure of glass fragility. [39] A striking aspect of the correlation of Scopigno and coworkers [38] is that it formally relates the fast dynamics scattering properties of materials in their glass state with the a parameter describing the change of the rate of relaxation and diffusion of the fluid in its liquid state above $T_g$. This observation complements other observations [35,40-44] indicating a direct link between the structural relaxation time $\tau_\alpha$ and $<u^2>$, which provide further evidence of links between the fast dynamics on the timescale ps with structural relaxation times that can be as large as minutes near $T_g$. In our view, these interrelations are remarkable given differences in the timescales involved.

It is often imagined by 'analogy' that the vibrational motion in cooled liquids is akin to that in perfect crystals at low temperatures where the atoms uniformly are localized in wells and that the amplitude of this vibrational motions increases in the amplitude in a uniform way for all the particles in the system. The limited amplitude of motion created by surrounding particles creates 'cage' in which the particles are localized. Of course, this naïve picture rapidly ceases to be true in heated crystals as non-linear excitations spontaneously arise from the inherently anharmonic interactions that arise between the thermally 'excited' particles (see discussion below). This is an extremely rich topic from a general theoretical standpoint, but we may get a *concrete sense* of the nature of this phenomenon in our metallic GF liquid by directly visualizing its dynamics to compare with the 'homogeneous' picture of material dynamics suggested by perfect crystals at low $T$. We show a snapshot in Fig. 1(a) of values of the localization scale defined by the mean square particle displacement $<u^2>$ at a characteristic time scale on the order of the fast $\beta$-relaxation time. The simulation data is for an Al-Sm metallic glass investigated extensively in previous work and the simulations were performed for $T = 450$ K. Our purposes for showing this image is just to give a qualitative sense of the nature of the material. We see that the local mobility, as quantified by $<u^2>$, which is a ps observable, is highly non-uniform in this cooled liquid. This is the normal state of affairs in liquids in the $T$ range in which we normally use them. Liquids are often more heterogeneous in a dynamical sense than we normally consider.

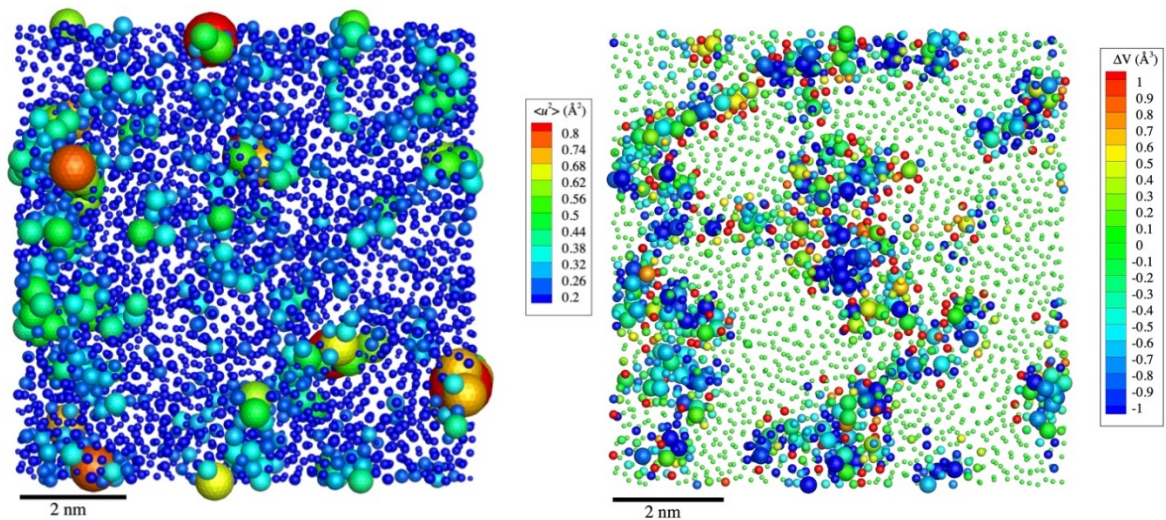

**Figure 1.** Local mobility and structural fluctuations. (a) Image of variable local $<u^2>$ values taken from a slice of 6 Å thickness metallic glass at $T = 450$ K. The color map to the side indicates the relative magnitudes of $<u^2>$ and the size of the particles also indicate the magnitude of $<u^2>$. (b) Image of variable local Voronoi volume values taken from the same slice in (a) $T = 450$ K. The color map to the side indicates the relative magnitudes of Voronoi volume. Large size particles indicate mobile atoms, median size particles represent those close to the mobile ones and the small size ones are the rest ones. We notice that there are three obvious populations of particles in this figure. There are relatively large regions (light green) in which the Voronoi cells have a volume close to the average for the entire material, there are complementary 'filamentary' regions in which there are well-packed particles surrounded by particles showing diminished packing efficiency. Despite this local density non-uniformity, the average density of the material changes smoothly with $T$. We discuss these local density heterogeneities further below.

Since the idea that there should be some 'structural' origin to these mobility fluctuations is apparently 'hard-wired' into the brains of material scientists, we also show for comparison the volume of the Voronoi cells associated with the atoms in our metallic glasses in Figure 1(b). While we see density is not completely decorrelated from mobility, the correlation between local density and mobility is weak, a point that we and others have made repeatedly in past work. [45-48] The variations in the local mobility simply cannot be explained as a "free volume" effects when we define this quantity in terms of local density. Of course, we can do much better if define a 'dynamic free volume' in terms of $<u^2>$ or the related quantity, positron volume, and we will discuss the value of these alternative free volume measures below.

The previous simulations of Betancourt et al. [28] have shown that the large values of $<u^2>$ seen in Fig. 1 predominantly have their origin in excitations in the GF liquid taking the form of particles moving collectively in the form of polymer-like structures, which we term 'stringlets'. The atoms in these excitations undergo almost ballistic or highly correlated displacements on a timescale on the order of a ps that are more reminiscent of particle motions in gases than particles crystals. [28], and there is direct evidence for this type of large-scale collective motion in dynamic neutron scattering measurements. [49]

Since the dynamics of GF liquids is evidently highly heterogenous even on a ps timescale, and we must push back to an even smaller timescales to comprehend the origin of this striking phenomenon. At the outset, we should acknowledge that the existence of this collective string like motion on ps timescales was discovered much earlier by several research groups. [50-54] We stress at the same time that these former authors did not offer much quantification about such structures, except to note that they grow progressively upon *heating,* which evidently makes these dynamic structures quite distinct from the dynamic string-like structures observed at long times in association with activated processes governing molecular diffusion and relaxation. [9,55-57] Zhang et al. were the first to quantify the length distribution and the temperature dependence of the 'stringlets' in the context of the glassy interfacial dynamics of Ni nanoparticles (See Fig. 3 of Ref. [58]). Later investigations by Betancourt for a polymer GF liquid [28] and our current metallic GF liquid (See below) are all qualitatively consistent with the early work of Zhang et al. and the pioneering qualitative studies of string-like collective motion on a ps timescale by Schober and others.

Betancourt et al. [28] also emphasized the importance of fast collective motion in influencing the amplitude $A_\beta$ of the $\beta$-relaxation process, as discussed above, but we shall see below that these string-like motions are not generally relaxational in nature. There are also collective normal mode motions having the same string-like structural form. In particular, the particles in some of these strings are undergoing *reversible motions*, except in rare instances when these localized modes induce particles to discretely 'jump' as consequence of their intrinsic large amplitude motions (and presumably large local stresses) arising from energy localization by these collective excitations. This phenomenon will be discussed below.

The observed dichotomy of string-like structures involved in both the relaxation and elastic response of the material in the fast dynamics regime deserves some comment. Certainly, we had not initially anticipated this behavior, and the possibility of the reversible motion never occurred to us before. With the advantage of hindsight, and recent observations on liquid dynamics emphasizing an energy landscape perspective [11], we can readily rationalize this behavior. A normal mode analysis of GF liquids indicates after an inherent structure quench to zero temperature generally reveals the existence of localized harmonic modes that have been of great interest recent in relation to understanding how the intrinsic disorder of liquids influences their normal modes in relation to crystalline materials. [59] Now if one takes the same fluid and performs an instantaneous normal mode analysis at a *finite temperature*, the localized vibrational modes separate out into two distinct populations of modes: localized stable and localized unstable modes. [60,61] The stringlets that we observe appear to be the physical manifestation of this mode bifurcation: localized stable and unstable modes (localization is implied by their highly limited spatial extent) so that it is natural that the unstable modes should be involved in relaxation processes involving fast particle intra-basin displacements, while others stringlets undergo the oscillatory motion required of stable modes and these companion modes contribute to the elastic response of the material. Discriminating these structures requires watching them for a timescale longer than a ps, something that we failed to appreciate earlier.

We show below that the average rate of large-scale jumping is entirely consistent with the Johari-Goldstein relaxation process, and the intermittency of this particle hopping process quantitatively explains the characteristic time dependence of the JG $\beta$-relaxation process. We discuss the string-like vibrational modes below in connection with their scattering signature, the Boson peak, but first we consider some dynamic heterogeneity aspects of the Johari-Goldstein $\beta$-relaxation process, the 'not quite so fast' relaxation process. We should mention that the possible importance of reversible string-like motion on a ps timescale for understanding the Johari-Goldstein $\beta$-relaxation was suggested earlier in a series of papers by Samwer and coworkers. [62-64] and we confirm below that this type of collective conspicuously arises in connection with particle jump events whose frequency of occurrence apparently determines the Johari-Goldstein $\beta$-relaxation time.

## B.  Mobile Particle Clusters, Intermittent Hopping and the JG-Relaxation Process

As just mentioned, the other 'fast' relaxation process is the JG $\beta$-relaxation process, which is the relaxation process of primary significance in glass materials. This situation arises because the $\alpha$-relaxation process is just too slow to be operative under normal conditions in the glass state, apart from slow property changes that may arise as the material 'ages' over long timescales. In our previous study of our Al-Sm metallic glass, we observed that the relaxation time associated with the JG relaxation process $\tau_{JG}$, determined from simulations of stress relaxation on the same the metallic glass studied in the present paper and the findings validated by measurements on the same material [3], we found that $\tau_{JG}$ *quantitatively* coincided with the lifetime of a particular type of dynamic heterogeneity in the GF liquid. In particular, it was found that $\tau_{JG}$ corresponds to the lifetime of the *mobile particle clusters* that generally exist in regions of the GF liquids in which there is relatively inefficient molecular packing and frustrated intermolecular interactions. In crude terms, these regions in the liquid, which appear to arise quite generally, are analogous to grain boundaries in polycrystalline materials, which are likewise regions of rich in mobile particles in the same technical sense as the GF liquids. [65] This picture of the physical origin of the $\tau_{JG}$ relaxation process conforms closely with the conceptual picture of the $\tau_{JG}$ relaxation sketched in the pioneering work on this relaxation process by Johari, who in turn attribute this general physical picture of the intrinsic nature of glassy materials to even earlier work by Tammann [66] [This historical background to modeling the JG $\beta$-relaxation process is discussed by Johari [66](See Page 324)]. We view the observations of Johari to be highly prescient so that these historical observations should be of some general interest.

The relationship between the JG $\beta$-relaxation time $\tau_{JG}$ and the lifetime of the mobile particles for our Al-Sm metallic glass is shown in Fig. 2. We note that Yu and coworkers [3,67] identified $\tau_{JG}$ with the lifetime of atom clusters undergoing string-like collective motion, $\tau_{Yu}$. This identification was based on direct comparison of experimental data for $\tau_{JG}$ obtained from dynamical mechanical measurements and simulations of shear stress relaxation in the metallic system studied in the present paper [3] and we show a comparison of these computational and experimental estimates of $\tau_{JG}$ with $\tau_{Yu}$ in Fig. 2. An earlier by Yu and coworkers[67] showed the correspondence between computational estimates of $\tau_{JG}$ with $\tau_{Yu}$ in a $Ni_{80}P_{20}$ metallic glass, a material exhibiting a very different pattern of glass-formation. The correspondence between $\tau_{JG}$ with $\tau_{Yu}$ is indeed striking and seems to offer an important clue into the intrinsic nature of the JG relaxation process in terms of a kind of dynamic heterogeneity. However, we emphasize here that while it is true that collective motion is an attribute of particles within the mobile particle clusters defined by Yu and coworkers[3,67], these clusters *cannot* be identified with the string-like clusters involving collective particle exchange motion on long timescales that are realizations of diffusive barrier crossing events in the fluid. [9,42,47,68-70] To clarify this point, we include the lifetime of the clusters involved in string-like collective exchange motion in Fig. 2, where we see that the relaxation time $\tau_{String}$, the string lifetime, is different from both the $\alpha$-relaxation time $\tau_\alpha$ and $\tau_{JG}$. This difference in timescale for the string lifetime results arises from a different definition of the strings. The clusters defined by Yu and coworkers are essentially equivalent to clusters that we have previously termed 'mobile particle clusters', which are super*clusters* of strings [9,36,57], structures of primary significance for understanding changes of the activation energy for diffusion and relaxation in cooled liquids. [9,42,47,68-70] We discuss the relation between the string clusters and mobile particle clusters in Supplementary Information, but *qualitative difference* in these definitions is evident from the difference in the average size of these mobile particle clusters at low temperatures. The mobile particle clusters exhibit significant branching, in general, and their average mass is far too large to be identified with the change of activation energy in cooled liquids. [9,36] Here we make this point by comparing the relaxation time $\tau_{Yu}$ of the mobile particle clusters of Yu et al., which they designate as the "string relaxation time", with the mobile particle relaxation time $\tau_M$ determined by a procedure that is now standard from previous work. [36,71] We discuss the determination of the strings and their lifetime in our companion paper where we show that these structures allow for quantitative interpretation of the $T$ dependence of the activation free energy in our Al-Sm metallic glass system. We should note that this has been a point of confusion in the past where other groups have likewise designated the mobile particles as being 'strings'.[72-75] These different classes of mobile particles should be carefully discriminated since they apparently have rather different relationship to the dynamics of the material. The present work, and the former work of Yu et al., are the first works to establish the significance of the mobile particle clusters on the dynamics of GF liquids.

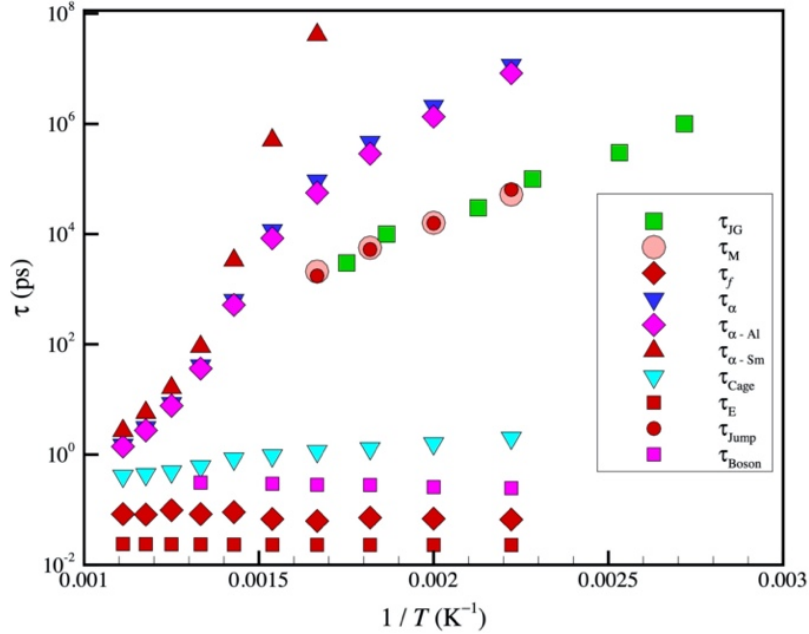

**Figure 2**. 'Fast' (lower band of relaxation times) and long-time relaxation times (upper band of relaxation times) of $Al_{90}$ $Sm_{10}$ metallic GF liquid. The legend indicates the primary $\alpha$-structural relaxation time $\tau_\alpha$ (as well as $\tau_{\alpha\text{-}Al}$ for Al and $\tau_{\alpha\text{-}Sm}$ for Sm), the fast-relaxation time $\tau_f$ describing the decay of $\beta$-relaxation of the intermediate scattering function, and the 'caging time' $\tau_{Cage}$ indicates the time at which the $\alpha$-structural relaxation process and the time at which the Debye-Waller factor $<u^2>$ describing the mean cage size is defined. $\tau_{Cage}$ in molecular fluids is typically on the order of 1 ps. The Boson peak time $\tau_{Boson}$ is defined as, $\tau_{Boson} \equiv 1/ (2\ \pi\ \omega_B)$, where $\omega_B$ is the Boson peak frequency, which is calculated below. $\tau_E$ is a measure of the average collision rate, which has a typical magnitude on the order of magnitude of $10^{-14}$ s. This 'phonon' relaxation time is the reciprocal of $\tau_E$ is the mean small-oscillation frequency of the atom in its potential energy well. [76] The Johari-Goldstein $\beta$-relaxation time $\tau_{JG}$, determined from MD simulations of the frequency dependent shear modulus of our Al-Sm GF liquid, was found to correspond to the lifetime of the mobile particles $\tau_M$ in our companion paper, and in the present work we show that $D/\ T$ strongly correlates with the inverse of the average mobile jump rate $\tau_{Jump}$, extending our observation in our previous work [10] that so that $D/\ T$ scales inversely to $\tau_{JG}$. Together these results provide a new interpretation of physical significance of $\tau_{JG}$.

The empirical correlation between the lifetime of the mobile clusters $\tau_M$ and $\tau_{JG}$ does not really explain the physical origin of this relationship. The evident physical significance of the lifetime of the mobile particle clusters means that we need to consider the specific underlying dynamical process by which the mobile particle clusters disintegrate. Since cooled liquids are inherently dynamically heterogeneous even on a ps timescales, it makes sense to look at how the properties of the previously identified fast particles and the remaining particles evolve over time to see if we identify any conspicuous difference in the dynamics of particles in these distinct mobility states that might offer a clue about how the mobile particles reorganize over time.

In Figure 3(a), we show the time evolution of the mean square amplitude of atomic motion $<u^2>$ excited by thermal energy at $T$ = 450 K (green), the local coordination number (CN) (blue) by summing up the number of faces of the Voronoi cell, reflecting fluctuations in the number of atoms around the representative atom and the fluctuations in the local potential energy felt by the test particle. We see that there is nothing at all exciting here- just fluctuations that probably take the form of white noise so that the fluid appears effectively *dynamically homogeneous.* Nine times out of ten this is what you would see, but if you happen to examine a mobile particle giving rise to the large $<u^2>$ noted above, then you see a very different evolution of the local properties. We show the corresponding evolution of one of these 'excited' particles in Figure 3(b). It is clear from this time evolution that the mobile particles are in a qualitatively different state. There is nothing subtle about this difference in dynamical state. Note the huge spikes in particle displacement reminiscent of earthquake intensity data and the potential energy is undergoing quasi-periodic jumps up and down. Correspondingly, we may expect large local thermal energy fluctuations as emphasized by Zylberg et al. [77], but we do not study these fluctuations here. In a previous study of these mobile particles in the context of the interfacial dynamics of crystalline Ni nanoparticles it was found that the potential energy jumps are *huge*- almost an eV in magnitude, but in the present bulk Al-Sm material the magnitude of these energy jumps are much smaller magnitude, typical having values around 0.1 eV. The local environment of the particles is also fluctuating in a dramatic way that tracks the potential energy changes. The amplitude of the energetic fluctuations is much larger for the excited mobile particles. Note that the large apparent deviation from equipartition of energy implied by these observations has been argued to be a general and neglected feature. [78]

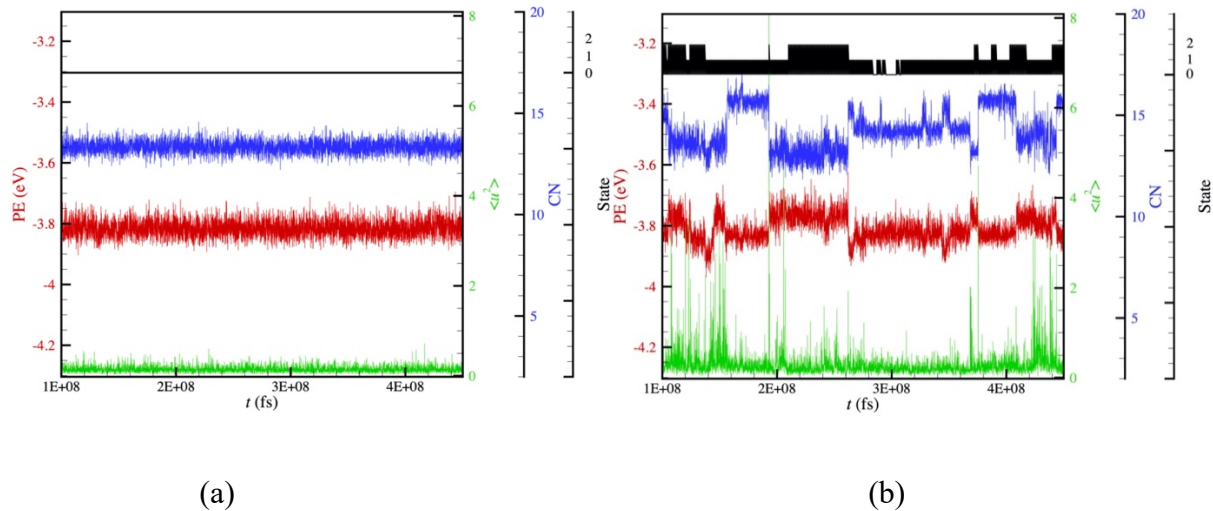

(a)                                                                 (b)

**Figure 3.** Mobility, coordination number and potential energy fluctuations of mobile and normal particles. (a) Time evolution of a 'normal' particle as determined by $<u^2>$, the local particle coordination number, and local magnitude of the potential energy felt by the particle (PE is red, $<u^2>$ is green, and CN is blue.) (b) Time evolution of a 'mobile' or 'excited' particle as determined by $<u^2>$, the local particle coordination number, and local magnitude of the potential energy felt by the particle. The black time series indicates the state designation, i.e., 2 – 'mobile' particles, 0 - 'normal', i.e., particles that are far away from mobile particles.

We see that the particles in the mobile particle state jump between different reasonably well-defined potential energy states and we define these states by its local position relatively to mobile particles, i.e., 2 = 'mobile', 1 = 'activated' and 0 = 'normal', and by applying this threshold we obtain the discrete jump time series shown at the top of Figure 3. Note that the jumps in dynamical state occur intermittently and below further quantify this phenomenon, which directly related to $\tau_{JG}$. First, we examine the strikingly large jumps in the mean particle displacements on a ps timescale, since these clearly dominate the average value of $<u^2>$ even though there are not that many particles in this excited state. Before proceeding, we note that telegraph-like signal time series of jumps in Fig. 4 persists up to a time on the order of the lifetime of the mobile particle clusters, which can be a very long time in cooled liquids. [10] Over much longer timescales the particles ultimately exchange between these rather different dynamical states.

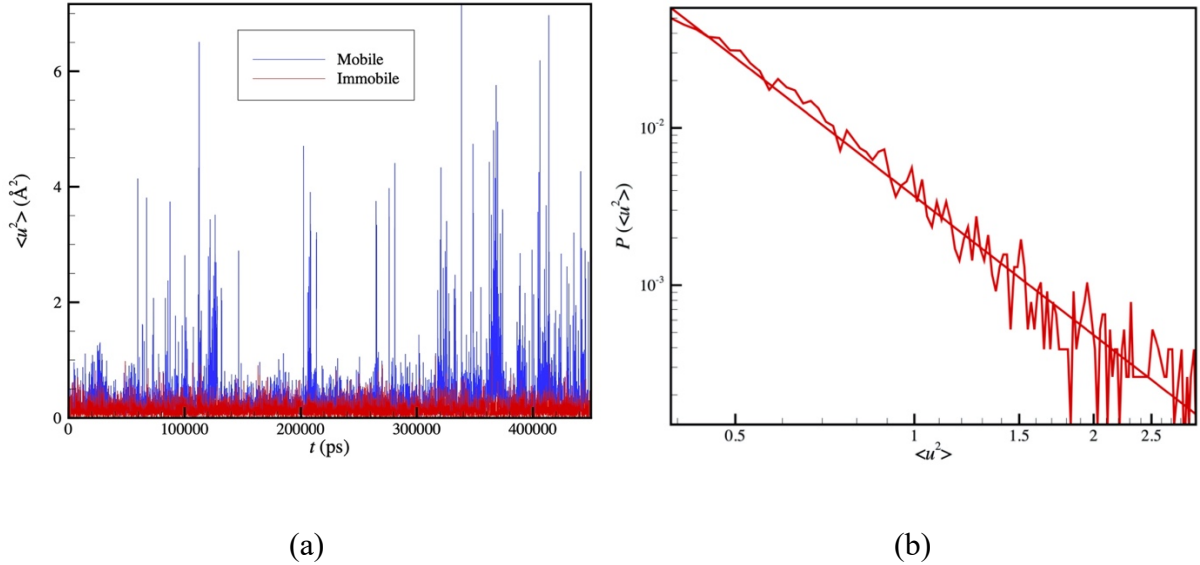

(a)                                                                  (b)

**Figure 4.** (a) Time evolution $<u^2>$ for mobile and normal particle in AlSm metallic GF liquid at $T$ = 450 K. (b) Distribution function of the height ($<u^2>$) of quake-like displacements in (a).

We directly compare the jumps in $<u^2>$ for the representative mobile (blue) and normal (red) particles in Fig. 4(a). There is a large difference in the relative displacements as described in an earlier work on a model polymeric GF liquid [28], the interfacial dynamics of Ni nanoparticles [58], the interfacial dynamics of ice [79] and the internal dynamics of the protein ubiquitin. [46] The phenomenon would appear to be *universal* in condensed materials when anharmonic interactions, which translates into almost any form of condensed matter from perfect crystals at low temperatures, i.e., most all real materials. If we further quantify the size distribution of these jumps in $<u^2>$, then see the data in Fig. 4(b) exhibits a power-law distribution as found in the intensities of earthquake data [58] and if we examine this scaling for other temperatures, we find the magnitude of the scaling exponent increases upon cooling, as illustrated in Fig. 5 and its inset. This is a behavior that we have repeatedly observed in the broad class of condensed materials systems noted above. We next quantify the rate of jumping in the mobile particle dynamics.

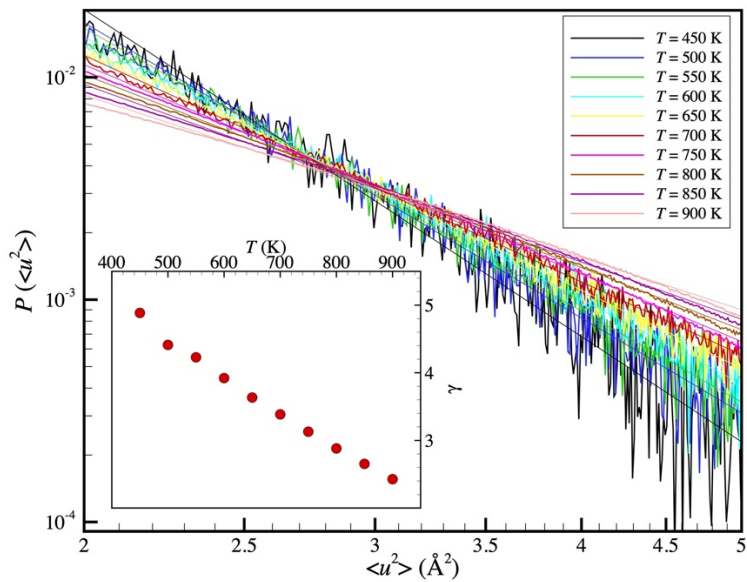

**Figure 5.** Distribution function of the height ($<u^2>$) of quake-like displacements of mobile particles at different $T$. Inset shows the temperature dependence of the scaling exponent, $P(<u^2>) \sim (<u^2>)^{-\gamma}$.

## C. Structural Origin of Potential Energy Jumps?

The search for a structural origin for mobility fluctuations in liquids has largely been a disappointing scientific enterprise[48,57,80], but there is a never-ending quest to find relations of this kind because of the innate belief in structure-property relationships by physical scientists. We mentioned in the introduction that the highly intuitive 'free volume' theory relating local mobility to local density has proven to be a failure. [48] We take some hope in the quest for a better structural understanding of mobility fluctuations from the time series in Fig. 3, however. It is evident that the mobile particles are jumping between 3 reasonably distinct potential energy states corresponding to well defined mobility states – 'activated', 'normal', and 'mobile'. You can't help but wonder if these potential changes reflect some kind of significant change in their local atomic environment. We give a cursory inspection of this possibility in this section based on ideas originally introduced by Rahman [81,82] that fluctuations in the local environment about particles in the liquid state might contain information about dynamics that are 'washed out' when spherically averaging the distribution of surrounding atoms as in the calculation of the pair correlation function and by ensemble averaging over the large ensemble of particles in the fluid. This averaging is particularly problematic for identifying structural indicators for mobility because the mobile particles are just a small subset of the particle population.

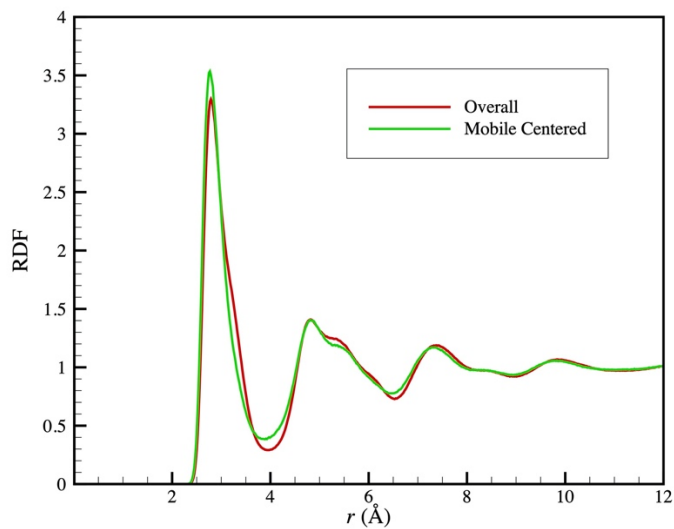

**Figure 6.** Pair correlation function $g(r)$ of mobile particle and surround particles compared to $g(r)$ of the fluid as whole.

We first look at the possibility that there is anything 'special' about the pair correlation function of a 'mobile' particle versus a 'typical' particle by calculating $g(r)$ for mobile particles and the fluid as whole. A comparison of these quantities in Fig. 6 shows that the first peak of $g(r)$ is a little smaller for the mobile particles and there perhaps a stronger splitting of the second peak, suggesting that there is some heterogeneity in the structure around the mobile particle. There is apparently no real striking change in structure that can be discerned from $g(r)$. In particular,  we do not see the sharp feature in $g(r)$ near its first minimum found by Laird and Schober [83] (See Fig. 3 of Ref. [83]) for mobile particles in a model soft sphere GF liquid that we had hoped to find as structural "signature" of this class of particles. Again, we find the structure dynamics relation to be elusive and we next consider a more refined too for examining local structure that does not involve the spherical averaging implicit in the calculation of $g(r)$.

To gain insight into how local structure varies in the system, it is useful to consider a Voronoi tessellation of the liquid defining a unique partition of space about all the particles at any instant of time and considering the volume of the 'neighborhood' about each particle, the Voronoi cell volume. There are numerous studies of this kind in the study of the dynamics of liquids that have sought in vain to find a correlation between Voronoi cell size and local mobility. We mention for example a rather thorough study of this kind for polymer GF that arrived at a rather negative conclusion regarding the density-based 'free volume theory'. [48] As a point of reference for the $Al_{90}Sm_{10}$ metallic GF liquid studied in the present paper, we show in Fig. 7(a) the ensemble averaged volume distribution of the Voronoi cells $P(V)$ for a particular temperature, $T = 450$ K. It is evident that the most probable Voronoi volume is pretty much the same for the mobile particles, the particles in the shells around mobile particles, which might have been different given the spitting of the second peak in Fig. 6, and for all the particles in the system. Interestingly, the distribution function of the mobile particles is unimodal, while this is not the case for the particles in the first shell of atoms about the mobile particle, suggesting the mobile particles is primary aluminum atoms. From this finding we infer that while the volume of the Voronoi cells may not be changing that much in relation to particle mobility, the shape of the cells might be. We then looked at the distribution of the sphericity $S$ of the Voronoi cells, a well-known dimensionless measure of shape defined to equal 1 and to have a smaller value for another shape because of variational principle indicating that of all objects of a given volume the sphere has the minimum surface area. In particular, $S$ is defined as $S = \pi^{1/3}(6V)^{2/3}/A$, where $V$ and $A$ are the volume and area of Voronoi cell, respectively. The distributions of the Voronoi cell sphericity for the mobile particles and the other particles, including the particles in the first shell about the mobile particles is evidently distinct. The most probably shape of the Voronoi cells corresponding to the mobile particles is clearly more distorted than the other particles.

This shape distortion of the Voronoi cells can also be understood in terms of how many particles surround a given particle, the local coordination number (CN). An unsaturated layer of atoms around a particle, corresponding to the presence of a hole, or having an extra atom jammed in about the central particle relative to the average, clearly might have some ramifications for the particle to escape its local environment and this might conceivably impact mobility. We show the distribution of the CN in our Al-Sm metallic glass fluid at $T = 450$ K in Fig. 7(c) where we see that mobile particles tend to have a higher coordination number. This situation is very similar to water, another fluid undergoing FS glass-formation where the molecules exist in a more mobile state have a higher local coordination number. [84] This physical situation also bears some resemblance to a dumbbell interstitial defect in a crystal where an extra particle is 'jammed in' around another particle in the lattice, creating an unstable situation that makes these defects highly mobile, [85] and which gives rise to additional localized modes in the density of states of the material.[86] We will see below that there is direct evidence that this is also occurring in our model GF liquid. Granato [85,87,88] has developed a highly successful model of GF liquids based on the idea that liquids can be thought of as being crystals with a moderate concentration of dumbbell interstitials. This model clearly has some merit based on our observations, although we hesitate to call the excitations that we observe and explore below 'defects' because of their dynamic nature and lack of any structure related to a crystallographic symmetry. Excitation is a more neutral term.

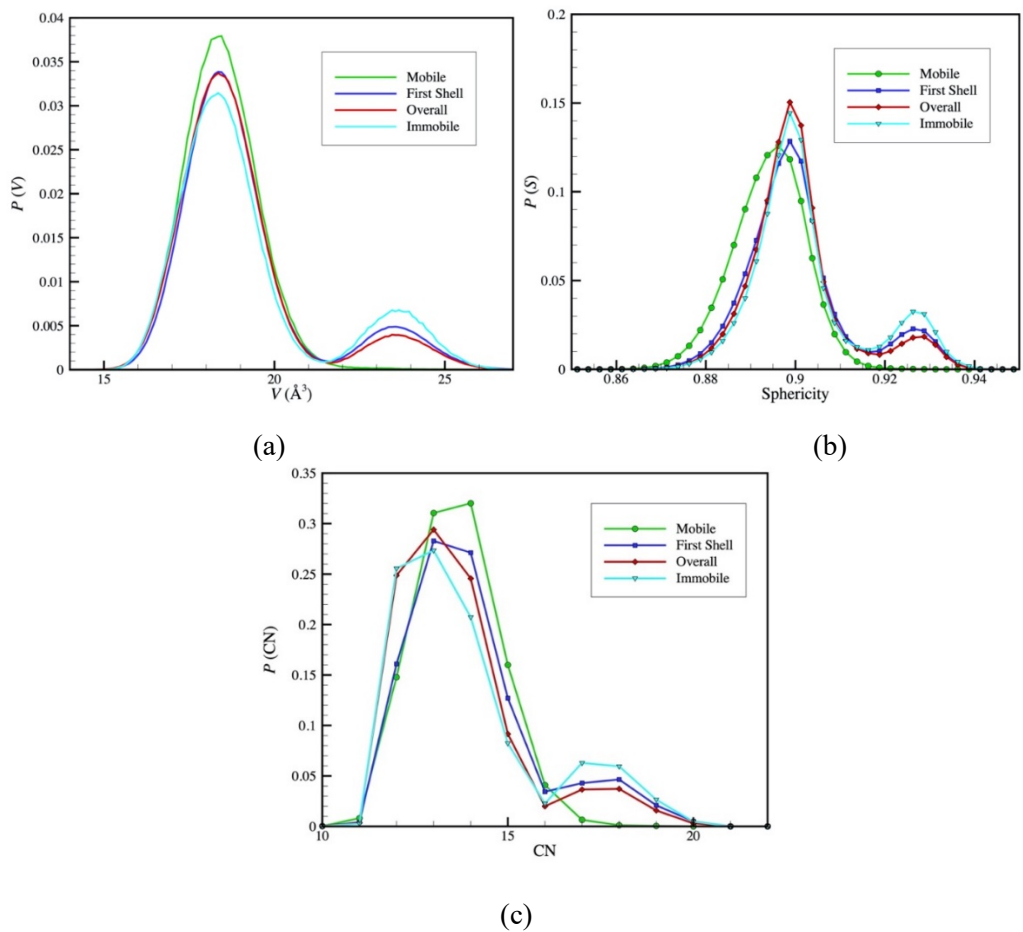

**Figure 7.** Distribution of Voronoi volume, sphericity and coordination number of Voronoi cells. (a) Volume size distribution, (b) sphericity distribution, and (c) CN of Voronoi cells of mobile particles, particles in the first shell of the mobile particles, immobile particles, and the fluid as a whole for $Al_{90}Sm_{10}$ metallic GF liquid.

A change of local coordination provides a clear rationale for the change in the local potential energy that we see in Fig. 3 where the jumps in the potential energy of the mobile particles occur in lock step to the jumps in the potential energy. It is apparent from the time series in Fig. 3 that there are *three* distinct coordination and potential states rather than the two we have so far identified. The other state can be accounted for by considering an addition form of mobility that we have not emphasized so far, the immobile particles. Recently modeling by Das and Douglas have recently defined distinct structural states in terms of coordination number in model two-dimensional fluids where each of these states corresponded to a distinct dynamical state [89,90]. A recent machine learning study of structure- mobility relations in glass identified local coordination number as a primary indicator of local mobility. [91]

The three potential states that we observe in our time series evidently are physically associated with changes in the local coordination number and thus the local potential of the atoms about the mobile particles. Clearly, the magnitude of these potential energy jumps can be expected to differ, depending on whether atoms are deep in the interior of a material or near its boundary or defect structures within it. The addition of additives that alters these jumps can be expected to modulate the magnitude of these jumps through their influence on the local cohesive interaction. This should greatly influence the rate of hopping as temperature is varied and we next turn to quantifying the rate of jumping and to its connection to the relaxation properties of our Al-Sm metallic glass melt.

The analysis of Voronoi volume and $g(r)$ considered averages over the entire material and it is apparent from Fig. 1b that the Voronoi volume non-trivial fluctuations in space in the form of extended domains separated from regions where the packing is uniform with a density close to the average for the medium as whole. We observe that that the characteristic size of the Voronoi volume fluctuations is ≈ 1 nm. The high density (low Voronoi volume domains) are presumably composed of bundles of chains of Voronoi cells having a near icosahedral form, which are surrounded by particles having relatively high Voronoi cell volume corresponding to the average of the material as a whole. Collective atomic motion is large concentrated in these 'grain boundary' regions of high packing frustration. Recent high-resolution imaging studies of metallic glass materials and molecular dynamics simulations (See Figs. 1 and 2 of Ref. [92]) accord remarkably well with the density non-uniformities shown in Fig. 1b, although the study by Feng et al. does not identify the close comingling of the high and low density packing within the regions occupied by the close packed atoms. We note that the emergence of large density fluctuations as in Fig. 1 b is not accompanied by a large change in the density of the change of the material as whole (See inset of Fig. 22 a of our companion paper [10]), although the thermal expansion coefficient changes it magnitude near $T_\lambda$. We tentatively interpret this spatial pattern of density non-uniformity to liquid-liquid phase separation of the metallic GF material. [93-96] (We discuss liquid-liquid phase separation in liquids in relation to fragile-strong glass-formation in our companion paper.[10]) It is not clear if the characteristic size of this pattern coarsens in time, as normally found in phase separation, or whether the pattern is pinned by the slow dynamics or the emergent elasticity of the material.

**Quantifying Colored Noise Associated with Potential Energy Fluctuations**

Fluctuations in potential energy are potentially measurable by spectroscopy and it is then of interest to quantify the potential fluctuations in Fig. 3. Fluctuations in the coordination number are expected to track these potential energy fluctuations so we confine our attention to potential energy fluctuations. We consider the power spectrum of the potential energy fluctuations of all atoms in our Al-Sm metallic GF material. The calculation of the power spectrum $S(\omega)$ of the potential energy fluctuation of the system $E(t)$ is defined by the transform,

$$S(\omega) = |\int E(t)e^{-2\pi i\omega t}dt|^2.\ [97]$$

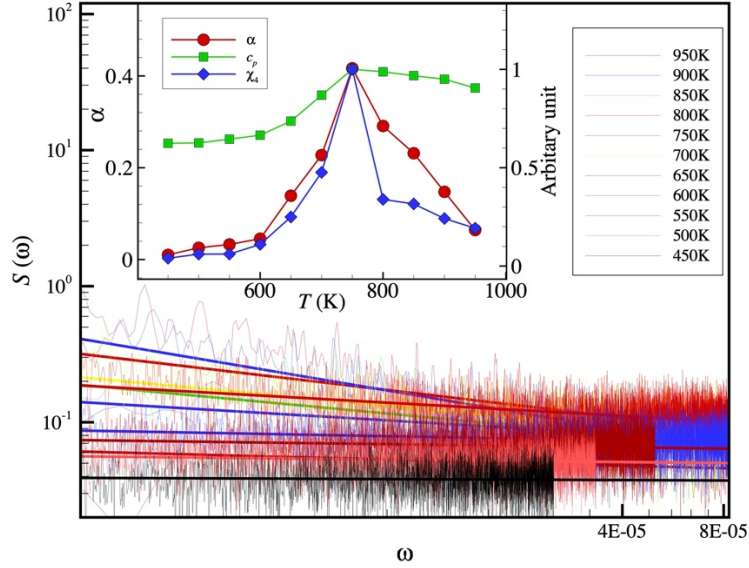

**Figure 8.** Power spectrum $S(\omega)$ of the potential energy fluctuations in a model Al-Sm GF liquid as a function of $T$. $S(\omega)$ shows an approximately power-law scaling with $\omega$, i.e., $S(\omega) \sim \omega^{-\alpha}$ and the inset to the figure shows the $T$ dependence of the colored noise power exponent, $\alpha$.

We observe that the colored noise exponent $\alpha$ describing the power law decay $S(\omega) \sim \omega^{-\alpha}$ varies non-monotonically with $T$, and this quantity peaking sharply at the lamda transition temperature $T_\lambda \approx 750$ K where the specific heat $C_p$ has a maximum (See our companion paper [10] for a discussion of this temperature) as an *equilibrium* feature in this GF liquid exhibiting a fragile-strong transition. [10] In the inset, we compare the $T$ dependence of $\alpha$ to the specific heat and the peak value of the 4-point density function, $\chi_4$. The estimation of these properties for our Al-Sm metallic GF liquid is described in our companion paper. [10] This variation of $\alpha$ with $T$ is very similar to the variation found in the potential energy fluctuations in the helix-coil transition of a polypeptides [98] and in the 'melting' of small clusters, [99] and is symptomatic of some kind of binding-unbinding transition. A very similar behavior has been observed in simulations of water where $c_p$, the peak height of $\chi_4$, the isothermal compressibility and the noise exponent for coordination number fluctuations all peak at the analog of $T_\lambda$ in our system. [7,8] We discuss this phenomenon at greater length in our previous work on the interfacial dynamics of nanoparticles where we also show that times series for $<u^2>$ also exhibits colored noise with a exponent that depends on $T$ that are of relevance for understanding mobility fluctuations in GF liquids. [100] We plan to report on this important phenomenon for the present Al-Sm GF material in a separate paper.

### D. Quantifying the Jump Rate in Relation to the Johari-Goldstein Relaxation Time

Since the jumps in the potential energy are well-defined and the lifetime of the mobile particle clusters $\tau_M$ is rather long at low $T$, we may examine the distribution of times between jumps (In general, one could discriminate between upward jumps in potential energy from downward jumps, but we do not pursue this refinement here for the sake of brevity.) Since we do not differentiate between upward or downward jumps in potential energy, we did not use potential energy as a criterion for determining the jump. The successful jump is defined based on displacement within certain time window. We may quantify this hopping rate more directly by simply putting a counter on these events $N_{jump}(t)$, just as we did in our initial study of the potential energy jumps in the context of the interfacial dynamics of Ni nanoparticles.[58] The average number of jumps $N_{jump}(t)$ as function of $t$ for different $T$ is shown in Fig. 9a, where the averaging is performed over the lifetime of different mobile clusters arising over the course of our simulation time. In Fig. 9(b), we normalize time by the lifetime of the mobile clusters $\tau_M$ and $N_{jump}(t)$ by the total number of jumps made during this time period, $N_{jump}(\tau_M)$ where see that all our data reduces to a convincing master curve. The lifetime of the mobile clusters $\tau_M$ clearly has significance in relation to the particle jumping process. In our previous study of the potential energy jump process, we found that the slope to be nearly linear, corresponding to a well-defined jump rate, and that the slope of this curve, the jump rate, was well-described by an Arrhenius $T$ dependence with a relatively large activation energy on the order of an eV, consistent with the JG $\beta$-relaxation process.[58] In the case of our Al-Sm metallic glass, the rate of jumping evidently decreases with increasing time, and the curvature of this jump rate curve obviously suggests a power-law scaling with $t$, i.e., the jumping is intermittent. This intermittency in the particle jumping process has been emphasized in many previous studies of GF liquids[101-104] so that the observation of intermittency in hopping is certainly not novel. The connection of this rate of jumping with $\tau_M$ is a new observation, however. We confirm our anticipation of a power-law in the jump rate by fitting to this functional form in Fig. 9(b). The fit is good (Note max deviation between data and fit curve) with an exponent of 0.4, but the time axis has been shifted by a constant so that time is given as $t - t_{shift}$. There appears to be a short induction time $t_{shift}$ before the power-law becomes established whose physical significance is presently unclear. We note that while the jumping process is an intermittent process, the power-law scaling exponent in this data over a large temperature range does not change with $T$ to within numerical uncertainty. Similarly, the probability of making a jump as a function of time $P(t)$ can be reduced to a near universal form by normalizing $t$ by $\tau_M$ and we show the result of this calculation in Fig. 10. The curve in this case is less universal, suggesting that we should separately consider the rate of jumping between the distinct states as in previous studies of the intensity of 'blinking' in quantum dots[105-110], a general, but physically unexplained phenomenon [108] that has obvious similarity to our observations on potential energy 'blinking' in glass-forming liquids. One of the general phenomenological observations is that the probability of a jump scales as power-law at long times with an exponent value that is normally near - 3/2, where this near universal exponent does not vary with $T$ [107], just we find for $P(t)$. We plan to consider these interconversion jumps between mobility states in greater detail in a future work. The mobile particles apparently disintegrate through the mechanism of these jumps, so we are led to ask what causes the jumps?

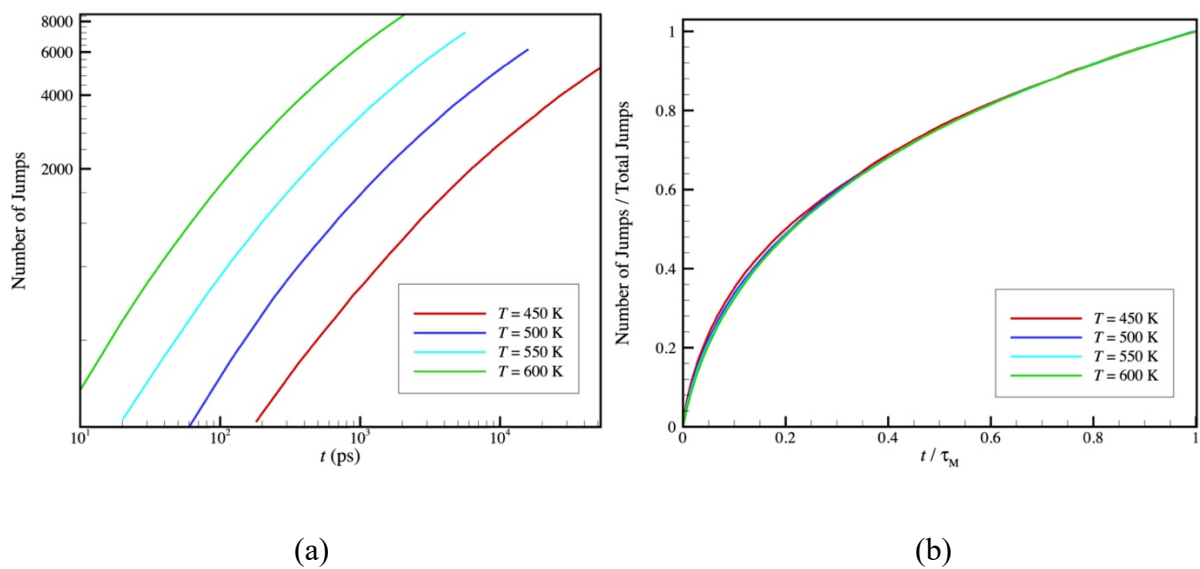

(a)                                   (b)

**Figure 9.** (a) Number of jumps of mobile particles during the lifetime of the mobile particle clusters $\tau_M$ for our Al-Sm metallic glass for various $T$. (b) The data is the same as (a), except that we have rescaled the x axis by $\tau_M$ and the y-axis has been normalized by the total number of jumps taken in the time, $\tau_M$. This data reduces to a convincing master curve.

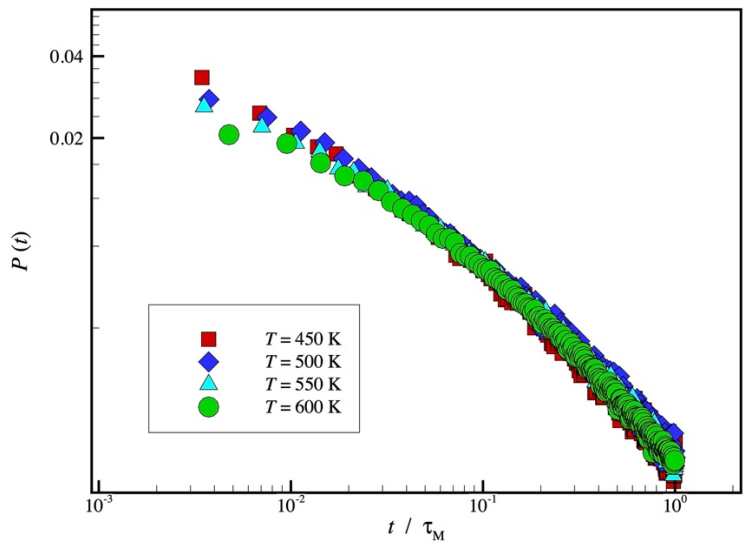

**Figure 10.** Distribution of average waiting times between jumps of mobile particles during the lifetime of the mobile particle clusters for our Al-Sm metallic glass for the various $T$ indicated where the time was scaled by $\tau_M$.

Apart from the significance of average waiting time for the mobile particles $\tau_{Jump}$ (obtained by fitting the probability of jump to a power-law distribution with an exponential cut-off) for better understanding the JG relaxation time, we may appreciate the relevance to molecular transport by comparing average diffusion coefficient $D$ in our metallic glass to $\tau_{Jump}$. Figure 11 shows that $D / T$ exhibits a power law scaling with $\tau_{Jump}$ with a decoupling exponent, $\zeta = 0.05$. This scaling, which means there is essentially no 'decoupling', strongly supports the interpretation of JG relaxation process as corresponding to an inter-basin hopping process on the energy landscape. [11] In our previous paper, we identified a numerous characteristic relaxation times of the Al-Sm GF material where the lifetime of the mobile particle clusters was the only relaxation time to exhibit a direct inverse relation with $D / T$. We have shown in the SI of our companion work [10] that average $D$ for the entire metallic GF material essentially coincides with the $D$ of the smaller atomic species Al, which in combination with the results in Figure 10 imply that the JG $\beta$-relaxation process is dominated by the diffusion of the smaller atoms, as emphasized in previous experimental studies. [63] We discuss this chemically specific aspects of the JG relaxation process in the Conclusion.

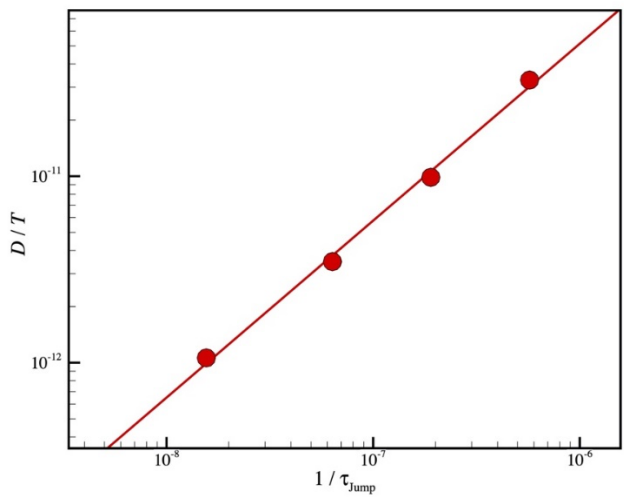

**Figure 11.** Relation between diffusion coefficient and average jump rime $\tau_{\text{Jump}}$, where is red line is the power law fitting. This inverse power scaling with a decoupling exponent scaling $\zeta = 0.05$ indicates that while landscape inter-basin hopping interpretation is related to $\tau_{\text{Jump}}$ we cannot directly identify this process with the JG $\beta$-relaxation process.

## E. Stringlet Modes and the Boson Peak

As mentioned above, previous work has identified string-like collective hopping motions on a ps timescale in model GF liquids [28,50-54], i.e., 'stringlets', and in the context of the interfacial dynamics of Ni nanoparticles [58], and since the potential energy jumps in time series for the mobile particles clearly occur on a timescale on this order, it natural to inquire whether the fast string-like collective motion actually *causes* these jumps. We also take this opportunity to more thoroughly investigate the nature of the motions of these peculiar thermal excitations. In particular, are they vibrational or relaxational collective modes? Schober suggested [50-52,111,112] that the strings correspond to quasi-localized modes reminiscent of dumbbell interstitials in heated crystals, which is certainly an interesting possibility, but previous simulation work has not demonstrated that the stringlets exhibit the oscillatory motion required by such an assignment. Rather, it is only established from previous work that these structures grow with heating, have a polymer-like structure and the atoms large amplitude motion in comparison with 'normal' particles in the material. It has also been observed on the basis of both experiment and simulation that the topology of the quasi-localized modes, which we call stringlets, can vary with the material. In particular, ring-like excitations are observed in some materials, exclusively linear in others and sometimes a coexistence of ring and string structures. In our Al-Sm metallic glass material, the stringlets are exclusively linear in their topology. [111] The stringlet modes exhibiting a collective rotational motion reminiscent of vortex suggest these vibrational modes are of a transverse mode nature while the stringlet modes having a worm-like character suggest a longitudinal mode and the mixed morphologies some sort of combination of these modes. The nature of these modes should be resolvable experimentally detectable by subjecting the material to incident radiation having different polarizations and measuring the extent of depolarization in the scattered radiation.[19]

We then identified a potential energy jump event that was shared by 4 adjacent mobile particles and it then became apparent by examining the motions involved in this collective energy potential jump event that we were observing a stringlet. The potential energy jump event indeed coincides with the potential energy jump, as we hypothesized. The correlated displacements involved with this stringlet is shown in Fig, 12. These atoms are moving in a concerted fashion on a ps timescale in the form of a string.

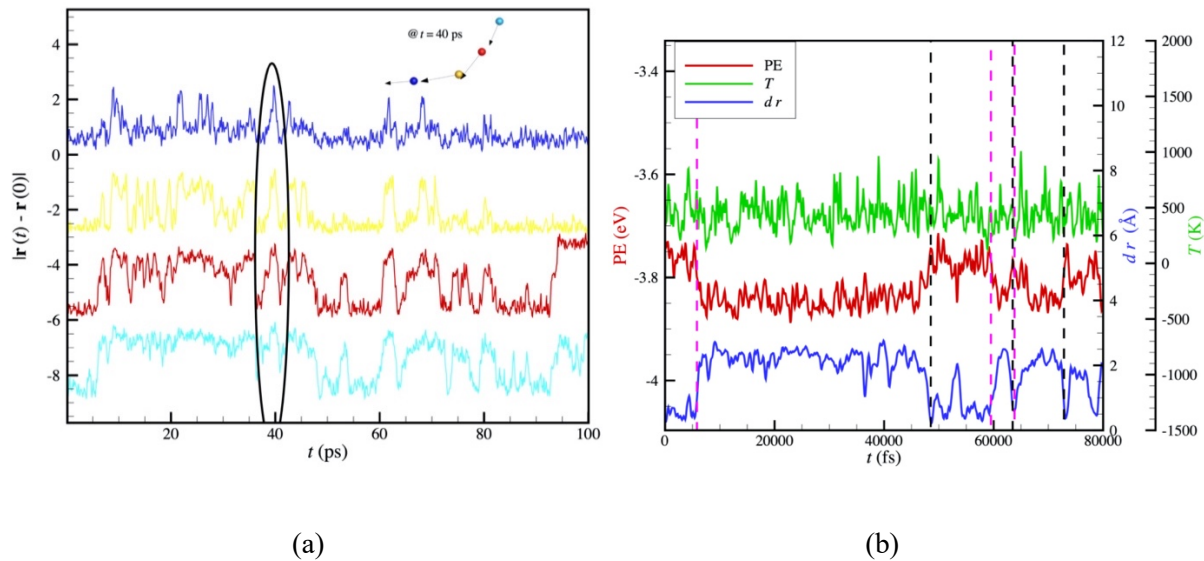

(a)

(b)

**Figure 12.** (a) Correlated jump event amongst mobile particle atoms. (b) Local potential, temperature, and displacement fluctuation for the cyan atom in (a) over time. The purple dashed line marks the start of one jump and the black dashed line marks the end of one jump.

There are many such events occurring amongst the mobile articles in our cooled liquid, and we show the size distribution of the stringlets in Fig. 13 where we find that the stringlets have a nearly exponential distribution of lengths, characteristic of an equilibrium polymerization process.[28] As observed in previous studies, [28,50-54,58] the stringlets grow upon *heating*, which is completely *opposite* to the strings in GF liquid which are associated within collective barrier crossing events. The strings persist to long time scales, comparable to the mobile particle clusters. It is important to discriminate these rather different string-like structures in relation to the dynamics of the collected liquid, despite their superficially similar geometrical form. Recent work by Hong et al. inferred the existence of a correlation length from Raman scattering measurements of the Boson peak (discussed below in Sect. G) that was interpreted as corresponding to a scale of cooperative motion on a ps time. [113] The characteristic scale $\xi_B$ of "dynamic heterogeneity" inferred from Boson peak measurements by Hong et al. was observed to grow slowly grow upon heating in the cases of model low and high molecular mass organic glass-formers: ortho-terphenyl and polyisobutylene. This trend makes perfect sense for the stringlets, but not for the strings. Although the identification with a scale of cooperative motion was intuitive by Hong et al. this exact interpretation was made by Granato [85] where the characteristic scale was identified with the length of the 'interstitialcy' localized mode forming a string of atoms undergoing oscillatory motion, an object consistent with our stringlets. The increase of $\lambda$ has a natural physical interpretation that we will discuss below. The growth of $\lambda$ corresponds to the softening of the stringlet mode upon heating and to a corresponding drop of the Boson peak energy towards zero.

We then examined the atomic motion of the atoms in the stringlets and noticed something that we should have anticipated before given the prior work by Schober identifying this type of fast dynamics collective motion with quasi-localized modes in the liquid. [50-52,111] The atomic motion of some of these clusters is *reversible*, except in infrequent cases when these motions lead to potential energy jumps, in which the particle displacements are nearly always irreversible, leading to diffusion. We illustrate the collective motion involved in the event shown in Fig. 12 (in a movie in SI.) It would then appear that stringlets correspond to a type of localized stable mode or 'excitation' that is topologically distinct from the phonons arising in prefect crystalline materials and these excitations provide a clear candidate for understanding the Boson peak, a universal feature found in scattering measurements on GF materials in the fast dynamics regime. This interpretation of the picosecond string-like collective motion as has been previously suggested and discussed at length by Schober and coworkers [111] This proposition may be checked by separately calculating the density states of both mobile particles undergoing stringlet motion and regular particles from their velocity autocorrelations functions, and then calculating the density of states of the entire material to determine whether the "extra" modes can be uniquely identified with the stringlets. We next consider such calculations for our model Al-Sm metallic GF material.

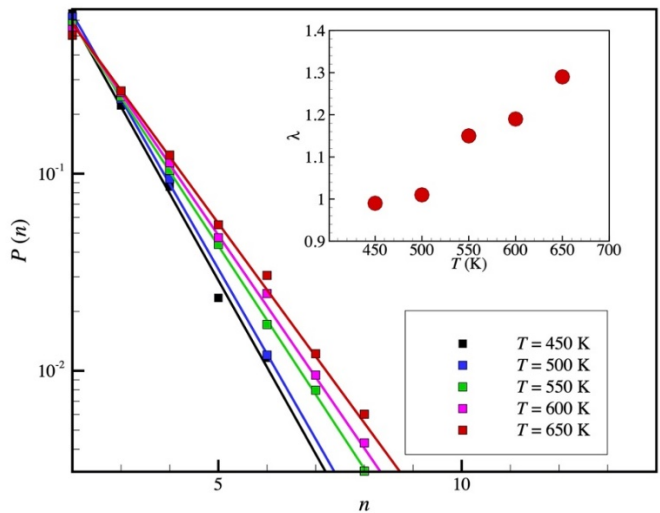

**Figure 13.** Distribution of lengths of stringlets lengths at different temperatures, as indicated, for Al-Sm metallic glass system, where the lines are the fitting of $P(n) \sim \exp(n / \lambda)$. Inset shows the temperature dependence of the average stringlet length $\lambda$, which evidently grows upon heating. This finding is consistent with the qualitive observations of Schober [50-52] and the later quantitative studies of the stringlets by Zhang et al. [58] in the context of interfacial dynamics of Ni nanoparticles and Betancourt et al. [28] for the dynamics of a model polymeric GF liquid. Note the stringlet length distribution is nearly exponential.

## G. "Smoking Gun" Connection Between Stringlets and the Boson Peak

As a first step in our calculation, we calculate the velocity autocorrelation function of one of the mobile particles involved in the evident collective motion in the event identified in Fig.12 and compare it to a 'normal' particle in Fig. 14 (both representative mobile and normal particles are Al atoms). The velocity autocorrelation function averaged over the whole fluid only shows some short time oscillations with local that rapid decorrelate thereafter, while the velocity autocorrelation function of the stringlet atom shows highly persistent oscillations, albeit of quasi-periodic nature. By contrast, the 'normal' atom shows oscillation (green curve) and thus no Boson peak. We interpret this as providing evidence for a chaotic breather mode [114], a spontaneously generated localized mode arising liquid from the anharmonicity of the intermolecular interactions and atomic granularity at an atomic scale which is perturbed from ideal oscillatory found in defective lattices by the effects of the intrinsic disorder of the liquid state and by interactions amongst these excitations as they compete to adsorb thermal energy and to consume each other.

We discuss evidence for the breather interpretation of these excitations in the context of the interfacial dynamics of Ni nanoparticles in a previous publication. [58] Since the notion of 'breather' or 'intrinsically localized modes' [115] are not universally familiar objects in the field of GF liquids we note the excellent description of an isolated excitation of this kind in a heated carbon nanotube where the mode can be readily visualized [116] (see Figs. 5 and 6 of Ref. [116]). Notice in this example that the amplitude grows in this mode progressively in time until the system "snaps", in this case generating a Stone-Wales defect in the carbon nanotube. In the liquid, we expect this type of excitation, albeit a chaotic breather, absorbs energy from its surround creating large local stresses until the system locally releases this energy in a burst of movement of the kind illustrated in Fig. 3(b) (The existence of these large local stress fluctuations in a GF liquid in the regime of non-Arrhenius dynamics was observed in pioneering early study of glass-formation by Chen et al., [117] but there has been no recent study quantifying has these fluctuations relate to dynamic heterogeneities within GF materials.) The density of (stable) modes $g(\omega)$ may be estimated by taking the cosine transform of the velocity autocorrelation function and we show this quantity normalized by the Debye model estimate for the density of states for simple phonons in three dimensions $g_D(\omega) \sim \omega^2$ to check for the nature of the normal modes associated with these particle motions and we show the result in the inset. The mobile particle clearly exhibits a peak in its reduced density of states, $g(\omega) / \omega^2$, while the normal particle does not. The origin of the Boson peak in our material is then evident.

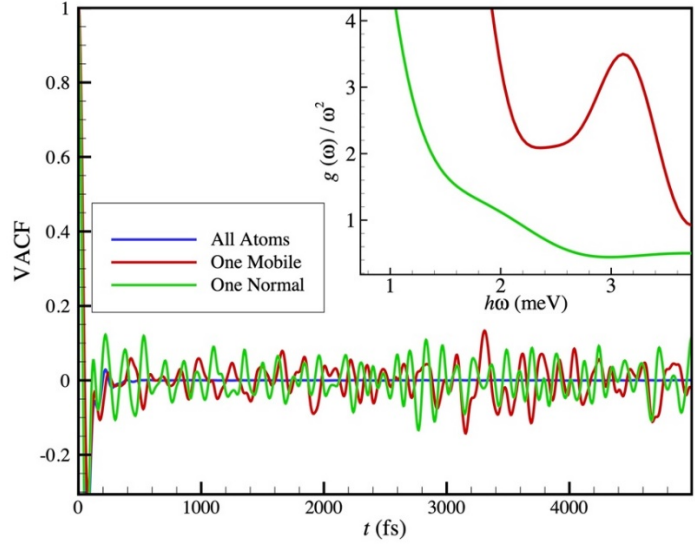

**Figure 14.** Comparison of velocity autocorrelation function for mobile particle, a typical normal particle and the average velocity autocorrelation function for all particles. Correspondingly, the mobile particle clearly exhibits a peak in its reduced density of states $g(\omega) / g_D(\omega)$, while the normal particle does not, as illustrated in the figure inset.

A calculation of the reduced density of states of the whole fluid obtained by averaging over all atoms is shown in Fig. 15 for comparison to the reduced density of states for a mobile and normal particle in Fig. 14. We see direct evidence of a Boson peak in our Al-Sm metallic glass and observe that Boson peak frequency $\omega_B$ varies in a range between 2 meV to 3 meV when $T$ is varied. This is a typical Boson peak as observed in inelastic neutron scattering or Raman scattering and corresponds to a characteristic time about a factor of 10 smaller than $\tau_f$ [1,118,119] and we find this to be the case in our own simulations. We include the Boson peak time for comparison in Fig. 2, based on the definition $\tau_{\text{Boson}} \equiv 1/(2\pi\omega_B)$. We tentatively interpret the fact that the Boson peak is not discernable in highly fragile GF liquids to imply the concentration of the stringlet excitations is lower in this class of fluids. We remind the reader that the Boson peak is a universal property of GF materials [120,121], but its intensity is not always large.

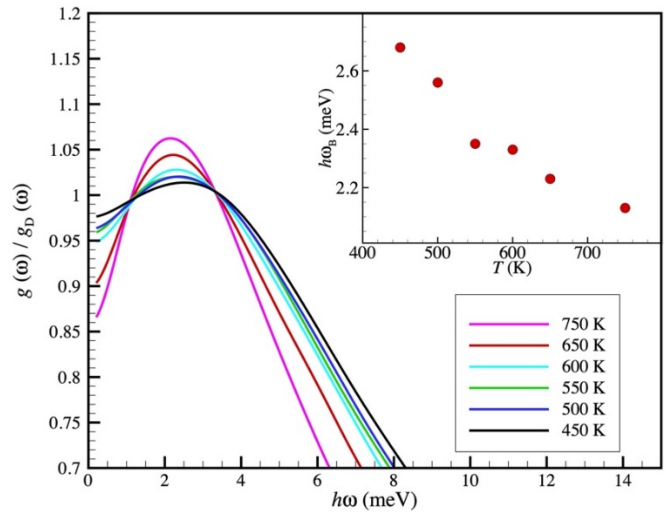

**Figure 15.** Reduced density of states $g(\omega) / g_D(\omega)$ for an Al-Sm metallic GF material over range of $T$. The inset shows the $T$ dependence of the Boson peak frequency, defined by the peaks in the curves. The decrease of the Boson peak frequency $\omega_B$ with increasing temperature means that the collective stinglet collective excitations decay more quickly at elevated $T$ and this mode 'softens' upon heating. The $T$ dependence of the Boson peak in the inset is typical of observations in real fluids, [122,123] although the data in the inset to this figure does not include the universal tendency for $\omega_B$ to plateau at low $T$.

As the Boson peak corresponds to the existence of a type of localized mode, albeit corresponding to a one-dimensional polymer-like structure, it is natural that the Boson peak will shift with a change of the shear modulus because of the well-known relation between the shear modulus to the speed of sound in the material. Consistent with this heuristic idea, Zaccone and coworkers [124] have recently predicted that the Boson peak frequency should scale specifically with the square root of $G$. In previous work we and others have shown that $G$ scales linearly with $k_B T$ / $<u^2>$ where $k_B T$ is the thermal energy. This seems to be an idea worth checking and accordingly we calculated the bulk $K$, shear modulus $G$ of our Al-Sm metallic glass over a wide range of $T$ because the wide practical interest in these mechanical properties of metallic glass materials. [125] We summarize our results in Fig. 16(a) where we observe both $K$ and $G$ to drop precipitously towards zero [126-128] upon approaching the high $T$ fluid regime of the metallic GF material (See ref. [10] our companion paper for a discussion of the transition between a high and low $T$ regime around 750 K in the Al-Sm metallic glass). This sharp drop in modulus is somewhat more rapid for $G$ so that the Poisson ratio $\nu$ [129] interrelating $G$ and $K$ drops increases in magnitude upon heating. Accordingly, we plot $\omega_B$ and $G$ versus $k_B T / <u^2>$ in Fig. 16(b) where we find that both of these properties scale nearly linearly with $\omega_B$ to a good approximation. We observed the same scaling of $\omega_B$ with $k_B T / <u^2>$ in a previous study of the Boson peak observed in association with the glassy interfacial dynamics of Ni nanoparticles. [58] [See Fig. 9 of Ref. [58]] This phenomenology seems broadly in line with the interpretation of the Boson peak as being a vibrational normal mode. Caponi has made this same point in a convincing way by showing that the variation of $\omega_B$ varies in concert with the Debye frequency of the material. [130] We note that this interpretation of the Boson peak as being due to chain like normal modes is consistent with a model introduced by Novikov [131] to describe the universal frequency dependence of $g(\omega) / g_D(\omega)$ for range near $\omega_B$. This model of the excess vibrational modes also naturally rationalizes the linear frequency dependence of the coupling constant $C(\omega)$ in the well-accepted Shuker-Gammon theory [132] describing the effect of the scattering of electromagnetic radiation from phonons, a key to interrelating density of states observations obtained from neutron and Raman scattering and between neutrons and specific heat measurements. Specifically, measurements have often confirmed this linear scaling, $C(\omega) \sim \omega$, to be nearly universal for $\omega > \omega_B$ [131,133-137] The universally observed near linear scaling of $C_p$ with $T$ at low temperatures, $C_p \sim T$ (In the $T$ range in which He$^4$ is a superfluid) can also be naturally rationalized from the existence of one dimensional chains of harmonic oscillators. [138,139] This linear dependence of the specific heat has often been seen in polymer materials[140] and carbon nanotubes [141,142] Given the results of the present paper, we think that the string model of the Boson peak of Novikov[131] deserves attention in the future.

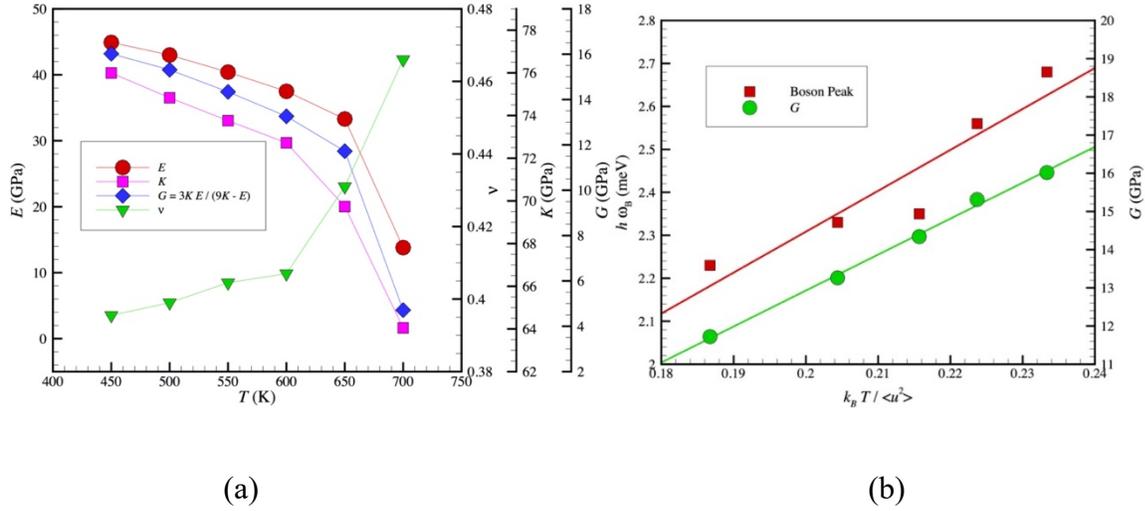

(a)                                                                (b)

**Figure 16.** Shear moduli, $<u^2>$ and the Boson peak and as a function of $T$ in an Al-Sm metallic GF material. (a) Bulk ($K$) and shear ($G$) modulus as a function of $T$. (b) Boson peak frequency $\omega_B$ and shear modulus plotted versus $k_B T / <u^2>$. A nearly linear variation of the Boson peak frequency and positron volume has been observed in polycarbonate GF liquids, [143] which a similar correlation given that the positron volume scales roughly linearly with $<u^2>$ for many GF liquids.[37,144]

## IV.    Conclusions and Discussion

We find that both the relaxational and vibrational dynamics in the fast dynamics regime are dominated at particle scale by string-like collective motion. In particular, the amplitude of the fast beta relaxation rapidly rises with heating in the glass state starting from near the Vogel-Fulcher-Tammann temperature $T_o$ [21-23] (See SI of our companion paper for an estimation of all the characteristic temperatures of our Al-Sm metallic glass) because of large scale, and highly localized collective motion, in the form of strings on a ps timescale, stringlets. Even though a relatively small fraction of particles is undergoing this correlated motion, the magnitude of the displacements is so large in comparison to "normal" particles they serve to dominant the magnitude of the Debye-Waller factor and thus the amplitudes of both the fast $\beta$- and $\alpha$-relaxation processes. We also see a reversible oscillatory mode in the fast dynamics regime having the same linear chain form and that Boson peak deduced from the atoms in these stringlet modes give rise to a Boson peak while "normal" particles do not. Thus, both the fast relaxation and elastic response are both caused by unstable and stable mode excitations taking a common form of string-like collective motion.

Relaxation beyond the ps regime also involves string like dynamics. The large scale string-like motion appears to drive large scale jumps in the atomic position, inter-basin displacements that we may associate with JG $\beta$-relaxation process, a relaxation process involving large scale hopping on the scale of the interparticle distance and relaxation time scale that is orders of magnitude longer than the fast $\beta$-relaxation time. There is also evidence of another fast dynamics process [145,146] involving a jump process related to intra-basin displacements on a scale much smaller than the interparticle distance. [11] We propose that this relaxation process is the same fast $\beta$-relaxation process that we study in the liquid regime, but that this relaxation process changes from being essentially *temperature independent* to a thermally activated process in the glass-state.[145,146] Interestingly, Luo et al. [147] have observed an abrupt change in the initial decay in the intermediate scattering function at a $T$ near, but below $T_g$, in which the relaxation time becomes Arrhenius with an impressively large change in the relaxation time prefactor for this relatively fast relaxation process. Interestingly, the super-diffusive displacement dynamics characteristic of the inertial dynamics of the fast relaxation process in the liquid regime above $T_g$, [See Eq. (2) of our companion paper], is preserved in the relaxation process so that "compressed exponential relaxation" is observed in the long time relaxation process normally accessible in macroscopic measurements. [147-151]

This emergent thermally activated relaxation process, which is highly prevalent in light scattering studies, thermodynamic measurements at very low temperatures and measurements of the "internal friction" of glassy materials, seems to be what experimentalists normally refer to the 'fast' relaxation process. [145,146] It is traditional to model this type of fast relaxation model by the asymmetric double well potential model (ADWP) of Gilroy and Phillips [152] in which some vaguely defined excitation is imagined to tunnel back and forth between two internal configurational states having different energies, hence the double wells. Oscillations between states of different energy is a basic to the existence stringlet excitations and the origin of this two-level characteristic has been investigated in the depth in the lattice localized mode analog of the stringlet excitation- the dumbbell interstitial. [85,86,153] Unfortunately, we do not see evidence of this activated form of the fast-$\beta$ relaxation in our simulations, presumably because we do not have data in the glass state. Notably, Dyre and Olson have introduced this same type of ADWP model as a phenomenological description for the JG relaxation process, but this work offers no physical of what the tunneling entity might be.[154] Here we suggest a clear interpretation of this emergent fast relaxation process as arising from a Gardner transition [155-157] in which the evolution of the material in its energy landscape becomes transiently trapped within minima with the metabasins which are readily traversed without thermal activation in the $T$ range above $T_g$ because of the relatively high kinetic energy of the particles.

Although the modeling and phenomenology of the fast $\beta$- and JG $\beta$-relaxation processes deep in the glass state resemble each other in many respects, making these relaxation processes easy to confuse, these basic relaxation processes in glass materials may be distinguished by a characteristic feature having its origin in the energy landscape motions of particles presumed to underlie these relaxation processes. We mentioned before that the fast $\beta$-relaxation process has its origin intra-basin motions or 'rattling motions' or that do not require thermal activation in the fluid regime, but the JG relaxation process involves intra-basin transitions over large energy barriers that separate metabasins. It is natural that the fast dynamics will ultimately transform to being activated in the glass regime as the thermal energy becomes insufficiently pass over these relatively small intrabasin barriers, resulting in the advent of activated transport, a phenomenon that we discuss further below. Importantly, we may expect the activation energies of the fast $\beta$-relaxation process and the JG $\beta$-relaxation process to be quite different. Consistent with this expectation, the activation energy of the experimentally estimated double-well potential for the fast relaxation process is far lower (by a factor on the order of 10) than the activation energy of JG $\beta$-relaxation in the extensive studies of fast relaxation by Rossler and coworkers.[145,146] We also point out recent observations of an additional $\beta$-relaxation in mechanical metallic GF materials that seems to occur quite generally whose activation energy is generally about ½ the value of the JG $\beta$-relaxation process. [158-160] We suggest that this mysterious '$\beta$'- relaxation' process can be identified as being the activated fast $\beta$-relaxation process just described. Together, the fast $\beta$-relaxation and JG $\beta$-relaxation processes are the primary relaxation processes of glass materials.

Although chemically specific effects on the fast dynamics of glasses is not the focus of the present paper, we make some comments on these chemically specific effects in view of the extreme practical importance and recent measurements aimed at comprehending this type of variation of the JG $\beta$-relaxation process with chemical composition. [62] Note that diffusion coefficient $D$ presented here is defined as the *composition weighted* (atomic fraction) average of the Al and Sm components (See Douglas et al. [35] for a discussion of a similar analysis for $D$ and the atomic component $D$ values for Cu-Zr metallic glasses having a range of compositions.) In the SI of our companion paper, [10] we quantify the $T$ dependence of the atomic diffusion coefficients ($D_{Al}$, $D_{Sm}$) over a wide $T$ range and their relation to $D$ for the present system. These observations indicate that the average $D$ for the metallic glass practically coincides with $D_{Al}$ and similarly $\tau_\alpha$ for the average structural time of the entire material was found in our companion paper [10] to be roughly equal the structural relaxation time of the Al component $\tau_\alpha$(Al). It is evident that the diffusion coefficient of the smaller atomic species (Al) can be orders of magnitude larger than the larger species (Sm), therefore, the average diffusion coefficient and structural relaxation time are both dominated by the fast-moving smaller atomic molecular species. This greatly accelerated dynamics of the smaller atomic species has been observed in radio isotope measurements on metallic glass materials.[161]

The direct relation between the average $D$ and the JG $\beta$-relaxation time that we have noted in our companion paper [10] also implies from the approximation $D \approx D_{Al}$ that the JG $\beta$-relaxation process should be correspondingly dominated by the rate of diffusion of the smaller atomic species. Measurement studies on model metallic glasses have correspondingly indicated a strong correlation between the diffusivity of the smaller atomic species and the JG $\beta$-relaxation time. [63] The extreme mobility anisotropy in the present material can be naturally understood physically from the rather large difference in the atomic size of Al and Sm. Al as an atomic radius of 1.25 Å while Sm has a radius 1.85 Å in terms of empirical atomic radii [162] so that Sm atoms are roughly over 50% larger than Al atoms, which is a size anisotropy that we may fairly characterize as being unusually large in magnitude. The strong dependence of $D$ on atomic species is well known in metallic GF materials [163] and an extensive review on this topic has been given by Faupel et al. [164] (See Sect. V), which also discuss various estimates of atomic size.

This discussion of chemically-specific aspects of the dynamics of the metallic glass system that we study brings us to finally consider why the JG $\beta$-relaxation process is so well separated from the $\alpha$-relaxation process in this material, a situation that normally allows to study the JG $\beta$-relaxation process without the strong overlap with the $\alpha$-relaxation process normally observed in metallic glasses. We suggest that the exceptionally large dynamic asymmetry that accompanies the large atomic size asymmetry is a contributing factor for this large timescale separation. We previously found in simulations of Cu-Zr metallic glass materials that degree of 'decoupling' between $D$ and the structural relaxation time correlated strongly with the fragility of glass-formation and relative mobility of the atomic species, as quantified by the Debye-Waller factor. [35] A large decoupling means their separation between the structural relaxation time and the characteristic hopping time associated with particle diffusion. The predominance of the average rate of atomic diffusion in the material by the smaller particles having higher mobility would seem to imply that smaller particles should give rise to a greater separation between the $\alpha$-relaxation process and the JG $\beta$-relaxation process, provided the JG $\beta$-relaxation process is generally controlled by atomic size and the effects of size on rate of atomic diffusion, as we may plausibly infer from our observations on our Al-Sm material. There is abundant evidence, however, that the strength of the JG $\beta$-relaxation process in both metallic and non-metallic materials is also dependent on the strength of the interparticle interaction, molecular stare of dispersion, nanoscale confinement, etc.[62,165-172] so we may not expect any simple geometrical interpretation of the strength of the JG $\beta$-relaxation process based on geometry alone. Despite the complexity of trying to understand the JG $\beta$-relaxation process in glassy broadly, it does appear that atomic or particle size generally is a relevant parameter for tuning the intensity of this relaxation process, along with the strength of molecular cohesion of particles with those in their surroundings. [62] Further work is needed to determine the separation in timescales between the JG $\beta$-relaxation process and $\alpha$-relaxation processes and the overall intensity of the JG $\beta$-relaxation process.

**Some Speculations about the Origin of the Activated Fast Relaxation Process**

The proposed change of the $\beta$-relaxation process from being inertial (depending mainly on kinetic energy) to activated dynamics in the glass state requires some explanation. In the liquid regime, *momentum fluctuations* in the fluid, associated with the fast collective motion and having their origin in the transport of momentum in the fluid through the motion of vortices, [173,174] makes the propagation of density fluctuations at the scale of the particle occur without any energy cost. This physics is well known responsible for the *propagation gap* in the dispersion relation for density fluctuation obtained from the $q$ dependence of the dispersion relation of the intermediate scattering function in a $q$-range around the size of the particles (See Montfrooij and Svensson [175] for a discussion.). As the $T$ is decreased and the entropy of the fluid corresponding drops to a low value, the propagation of density fluctuations necessarily becomes activated below some temperature at which an energy gap $\Delta$ emerges, exactly as in the emergence of the superfluid state in $^4$He. [176-178] Neutron scattering measurements of the dispersion relation for density fluctuations in metallic materials have indeed provided direct evidence for such a roton like energy gap in the dispersion relation obtained from the intermediate scattering function of a Zn-Mg metallic glass [179,180] and molten Pb [181], molten Na [182] and molten Ga [183], consistent with this suggestion. A convincing roton-like feature has also been observed in a Zr-Cu-Al metallic glass material [184]. We note that this roton feature in dispersion relation of GF materials and metallic melts has been greatly discussed in relation to understanding the low temperature dependence of the specific heat and thermal conductivity of glasses at low temperature. [185-188]. We reiterate a conjecture that we recently made in connection with a normal mode analysis of GF liquids [61] - we tentatively suggest that the Boson peak of GF liquids might be identified with the energy gap defining the excitation energy of the collective momentum fluctuations, an interpretation similar to the roton energy of liquid $^4$He. We believe that the results of the present work provide significant evidence supporting this interpretation of the Boson peak and we discuss some observational correspondences between Raman light scattering measurements on superfluid $^4$He and glasses to further support our case. Consistent with this discussion, recent simulations of metallic GF liquids [189] have indicated the occurrence of string-like collective atomic motion having a reversible nature in the form of closed vortex-like loops where the large vorticity in medium created by these anharmonic motions contributes substantially to plastic deformation and these motions also identified with a fast $\beta$-relaxation process having an activation energy on the order of 0.3 eV, a value that is much lower than the Johari-Goldstein relaxation process, in accord with our suggestion above that the fast $\beta$-relaxation process should become an activated relaxation process in the glass state.

Raman measurements allow for the direct observation of the Boson peak in superfluid $^4$He from the shift of the frequency of the scattered radiation from the frequency of the incident radiation arising from the scattering of the incident radiation from the roton modes, the energy shift being directly related to the energy of the excitations in the fluid because the density of states is maximal in this region. In this way, one can observe the overdamping of the roton excitations as you approach the lamda transition of $^4$He from below. [190,191] The Raman scattering observations involve an averaging over the dynamic structure factor[192], and allows for a direct estimate of the energy of the collective excitation [190], although inelastic neutron values allow for a more accurate estimations of the dispersion relation as a function of $T$, albeit with much greater experimental effort. [177] The observations of Raman scattering from GF materials are remarkably similar to observations on superfluid $^4$He, except there is a large background frequency-dependent scattering from the Brillouin scattering deriving from the fast relaxation process overlapping the scattering response from the localized modes that requires subtraction.[193] We again see an overdamping of the non-phonon vibrational modes upon heating so that these modes no longer scatter light, just as in the Raman measurement on [4]He in its superfluid state.

There are also similarities with a popular molecular model of the structural form of the roton excitations in superfluid [4]He (The reader is warned that no general agreement exists regarding the molecular scale nature of the excitations responsible for the roton currently exists.) with the model of its proposed counterpart in GF liquids described in the present paper. In particular, Feynman [194,195] argued that the rotons of [4]He involve molecular vortices atoms moving permutationally in the forms of strings that form closed loops, the roton being the minimally-sized 'polymer' loop of this kind. Williams and others have elaborated this model [196-199] into a full-blown molecular theory of the superfluid transition and Montfrooij and Svensson [175] have discussed the merits of this model in interpreting their inelastic neutron measurements on [4]He. Despite the physical appeal of this model of the roton, there is still no general consensus on the physical nature of this excitation after many years of theoretical and experimental research. We also note the distinction that while the excitations we observe are polymer like in structure, grow upon heating [28] and involve atomic permutational motion, these structures do not generally form closed loops. It is also worth pointing out that an ideal Bose condensate does not exhibit superfluidity and that the collective behavior that we see in association emergent material rigidity is a classical molecular dynamics effect.

The basic picture that emerges from our study, which complements a former study on the fast dynamics of a polymeric GF liquid [28], is that the fast $\beta$-relaxation progressively "shuts down" as channel for relaxation in cooled liquids as molecular caging emerges, requiring activated mass and momentum diffusion that must occur on progressively longer timescales with cooling, accounting for the emergence of the $\alpha$-relaxation process. When the entropy of the fluid becomes critically small, regardless of whether it arises as a consequence of equilibrium or non-equilibrium conditions, the system undergoes a 'glass transition' in which fluid transforms to a state that the fast relaxation process acquires an activated rather than inertial character and excitations associated quasi-localized modes become a prominent feature of the material response in its 'glass' state. Recent measurements on metallic glass materials have indicated a sharp change in the fast relaxation process slightly below $T_g$, [147] where this process transitions to becoming thermally activated with an activation energy on the order of 0.1 eV, which is about a factor of 10 smaller than the Johari-Goldstein $\beta$-relaxation process and consistent with the fast relaxation process in the glass state discussed above. The extraordinary thing about this transition in the fast dynamics in these measurements is that the prefactor changes by a factor on the order of 10 orders of magnitude. There does seem to be evidence of the anticipated transition in the mature of the fast beta relaxation process as the material enters the glassy state. This type of transition in the fast dynamics from passing from the liquid tom glass states clearly requires further investigation.

This 'entropy catastrophe' is also characteristic of the formation of the superfluid state in $^4$He. [200,201] Further work is obviously required to conform this general picture and to further quantify the relaxation processes and different types of dynamic heterogeneity that are implicated the structural basis of the complex dynamics of this class of materials. It would be of obvious interest to better quantify the collective excitations in glasses, following similar methodologies, as used for many years for liquid $^4$He. We also note that the oscillating nature of the stringlet mode fits very well with heuristic two-level system models of the low $T$ properties of GF, providing a tangible physical picture of these excitations and we anticipate that we should be able to identify rational estimates of the energetic parameters, distribution of activation energies, etc. that have also be simply assumed in the popular asymmetric double-well potential (ADWP)[152] model and similar phenomenological models used to fit relaxation and thermodynamic data of glass materials. [145,146] Interestingly, these experimental studies of many glassy materials indicate that the noise exponent of fits to the ADWP model exhibits a peak near $T_g$ that seems similar to the peak of the noise exponent in Fig. 8 of the present paper. These observations merit a corresponding study of mobility fluctuations in our metallic GF material, as we have done for the interfacial dynamics of Ni nanoparticles [100], and crystalline Ni [202]and ice interfaces[79], to see if we can understand the fundamental cause of the non-monotonic variation of the noise exponent governing fast relaxation in glass materials.

## Acknowledgements

H.Z. and X.Y.W. gratefully acknowledge the support of the Natural Sciences and Engineering Research Council of Canada under the Discovery Grant Program (RGPIN-2017-03814) and Accelerator Supplements (RGPAS-2017- 507975).

## Data Availability Statements

The data that supports the findings of this study are available within the article [and its supplementary material].

## Supplementary Material

See supplementary material for the $T$ dependence of mobile particle and excitation concentrations, quantification of potential energy fluctuations, elastic modulus calculation, determination of the Einstein relaxation time or average collision time as function of $T$ from the velocity autocorrelation function, average cage size from first passage time analysis, comparison of VFT fitting to the observed variation of the structural relaxation time as function of $T$, and analysis of the non-Gaussian parameter to identify the Boson peak frequency. We also include a movie showing the cooperative motion associated with atoms giving rise to the Boson peak in our material.

# Supplementary Information:
# Fast Dynamics in a Model Metallic Glass-forming Material

Hao Zhang[1†], Xinyi Wang[1], Hai-Bin Yu[2], Jack F. Douglas[3†]

[1] Department of Chemical and Materials Engineering, University of Alberta, Edmonton, Alberta, Canada, T6G 1H9

[2] Wuhan National High Magnetic Field Center, Huazhong University of Science and Technology, Wuhan, Hubei, China, 430074

[3] Material Measurement Laboratory, Material Science and Engineering Division, National Institute of Standards and Technology, Maryland, USA, 20899

[†]Corresponding authors: hao.zhang@ualberta.ca; jack.douglas@nist.gov

## A. Simulation Sample Preparation and Cooling Rate

In a previous study, a Sm atom centred '3661' short range order (SRO) motif has been identified as an abundant SRO unit in $Al_{90}Sm_{10}$ metallic glass system using ab initio molecular dynamics simulation, similar to full icosahedral tessellation in $Cu_{64}Zr_{36}$ metallic glass [1]. More recently, molecular dynamics simulations were employed to investigate the dependence of this SRO motif on the cooling rates, where the liquid $Al_{90}Sm_{10}$ specimen was continuously cooled from 2000 K to 300 K with five different cooling rates, i.e., $10^{13}$, $10^{12}$, $10^{11}$, $10^{10}$, and $10^9$ K/s. [2] It was found the population of '3661' Sm centred motif increased sharply upon $T_g$ and then gradually approached to a constant value at low temperature region, and the total population of the '3661' motif was sensitive to the cooling rate, i.e., the lower the cooling rate, the higher the motif population. The behavior of '3661' SRO clusters is very similar to the behavior of icosahedral SRO in CuZr metallic glasses. However, such SRO motifs do not tend to form interpenetrating network even at the temperature well below $T_g$.

In the current study, in order to obtain a well equilibrated $Al_{90}Sm_{10}$ at low temperature, we applied an even lower cooling rate (one order lower than the slowest cooling rate used in the above study[2]), i.e., a cooling rate of $10^8$ K/s from 2000 K to 200 K. Although an extremely low cooling rate was employed in the current study, we expect that the system still would not reach complete equilibrium at low temperatures where the reciprocal cooling rate becomes longer than the relaxation time, which is a common difficulty with any low temperature study of GF liquids.

In our companion paper [3], we mainly characterize dynamic heterogeneity in this $Al_{90}Sm_{10}$ metallic glass that exhibit fragile-strong transition with focus on relating the relaxation time of JG $\beta$-relaxation and structural relaxation $\tau_\alpha$ to mobile cluster and immobile particle lifetime, quantitatively describing the diffusion and structural relaxation using string model of glass formation, and characterizing $\lambda$-thermodynamic transition. Some equilibrium and structural properties such as hyperuniformity, structural relaxation of different elements, which are sensitive to their chemical aspect of the metallic glass, are also examined and discussed. While in the current paper, we mainly focus on the physical nature of JG relaxation process, fast relaxation process and Boson peak, universal relaxation process in all types of GF materials and the types of dynamic heterogeneity associated with the dynamical processes.

## B.  Temperature Dependence of Mobile Particle and Excitation Concentrations

The concentration dependence of interstitials plays a central role in the Granato theory of glass-forming (GF) liquids and defective crystals and the strong dependence of the shear modulus on the concentration is central to this model. [4-6] Since we have found clear evidence of excitations in our metallic glass that are consistent with these structures being intrinsically localized modes, albeit not really perfectly equivalent to dumbbell interstitials, it is then of importance to quantify how the concentration of the excitations influence the properties of our metallic fluid, with some emphasis on the high frequency modulus or glassy modulus that governing the shear deformation response of our metallic glass material in its glassy state below the lambda transition temperature $T_\lambda$ = 750 K (See SI of paper I[3]) to check consistency with the Granato picture and to better understand this basic structure-property relationship, structure being used in relation to excitations who are best understood in the course of the dynamical evolution rather than static entities.

The computation of the concentration of the interstitials is a laborious process, even in the case of heated crystals [7,8], but the term 'arduous' comes to mind in the case of large non-crystalline systems, such as Al-Sm metallic glass-forming material. In the course of our study, we came across an interesting study that attempted to quantify the excitation concentration in high density amorphous ice using a combination of experimental, reverse Monte Carlo lattice dynamics, diffraction measurements and molecular dynamics simulations. [9] Given the strong parallelism between the dynamics of supercooled water and ice with the dynamics of our Al-Sm system, we think that this study makes for a particularly interesting comparison to our present work. Tse et al. [9] identified excitation structures in high density amorphous ice by calculating the velocity autocorrelation of all the atoms in their system where they noted that about 20 % of the molecules were in state that gave a large contribution to the low frequency density of states that was not of the usual phonon origin. It was also observed that these excited particles formed the same type of stringlet form of most open small polymeric chains of atoms as we have observed in our own simulations. This material probably has an exceptionally high concentration of these defects, as independent Rahman measurements have shown that the high-density form of amorphous ice exhibits a high density of "two-level systems", while the low density and less mobile form of amorphous ice, shows no detectable excitations of this kind. [10] We note that the study of Tse et al. [9] involved a combination of experimental and simulation methodologies that allowed it to come to the conclusions stated above regarding nature of the collective excitations of this material.

Now if we wish to replicate this for our own system, which is considerably larger, we would have to calculate the velocity autocorrelation function for all the particles after a large number of ps timescales and for a wide range of temperatures ($T$). This might be useful in the future, but we take a simplified approach to this problem to gain at least some qualitative insight into how the excitation might vary with temperature. We found before in our earlier work on dumbbell interstitials in heated bulk crystalline Ni [7] that the concentration increased with temperature in an Arrhenius fashion, and more importantly for the purposes of the present paper, the number of mobile particles in the fluid varied in direct proportion to the number of interstitials. In particular, each dumbbell interstitial defect created about 20 mobile particles defined by having a large

Debye-Waller factor $<u^2>$ in comparison with the background 'normal fluid. The activation enthalpy $\Delta H_i$ for forming these defects was very large, $O(1 \text{ eV})$, but defects of this kind arise in appreciable numbers because the activation entropy $\Delta S_i$ is corresponding also very large. The large magnitude of the activation free energy parameters have great significance for understanding the properties of metallic materials, given the large influence such defects have on the elastic and relaxational properties of these materials, as emphasized by Granato [4-6] and others [11-13] following his pioneering observations [14] and modeling [4-6] of the influence of interstitials on both the thermodynamic and dynamics properties of metallic materials. We note that the concentration of interstitials only became as large as 0.13 % near the melting temperature $T_m$, but these defects clearly had a large role on the dynamics of the heated material, serving to initiate string-like collective atomic motion, etc. We did not examine the influence of these defects on the shear modulus, but other authors [15,16] have shown a large impact of the dumbbell interstitials on the shear modulus and other properties of crystalline metallic materials, even when their concentration is so small. The explanations for why these defects are so "pregnant" for altering material properties is discussed by these authors and the reader is pointed to this very interesting work since we have nothing to add to their arguments.

Based on these prior observations, and accepting the general philosophy of the Granato theory [4-6] that there are well-defined thermal excitations exist in glass-forming liquids at low temperatures having some properties common to interstitials in crystals given that we actually observe such excitations in our simulations, we probe the concentration of these excitations by simply examining the temperature dependence of the mobile particles and assuming that the concentration of the mobile particles is about a factor of 20 times larger than the concentration of the thermal excitations, which we term "breathers" rather than defects because of these intrinsically localized modes give rise to a generic tendency for energy oscillations in both their crystalline and amorphous solid milieus in which they actively influence material properties. We show the temperature dependence of the fraction of atom in our Al-Sm metallic glass that are in a mobile particle state in Fig. S1 where we see that the variation of this quantity correlates strongly with the change of the fraction of mobile particles $\Phi$ moving collectively in string-like clusters relative to those remaining in their unassociated or isolated "monomer" state. In paper I [3], we discuss evidence the string-like clusters to temperature dependent changes in the activation energy for diffusion. We see that the mobile particle fraction is saturating to a high temperature limit value near 0.45 corresponding to an excitation concentration of about 2 % if we adopt the factor of 20 ratio between mobile and breather particles.

We may imagine the factor of 20 to arise from the first and some of the second shells of mobile particles directly surrounding the breather. The average number of atoms in the first shell is $\approx$ 14 and in first and second shells is $\approx$ 60. To check this simple idea that the breathers are creating a local distortion of the local dynamics, we examine the dynamics of the first shell around a breather in Fig. S2. The red time series shows the colored noise fluctuations of the breather ( $\sigma$ = 0.64) and the blue time series correspond to the twelve atoms about the breather. The breather is clearly exciting the atoms in its nearest-neighbor shell into a state of large amplitude motion, i.e., mobility. The excitations in the fluid appear to be "enslaving" those in their environment.

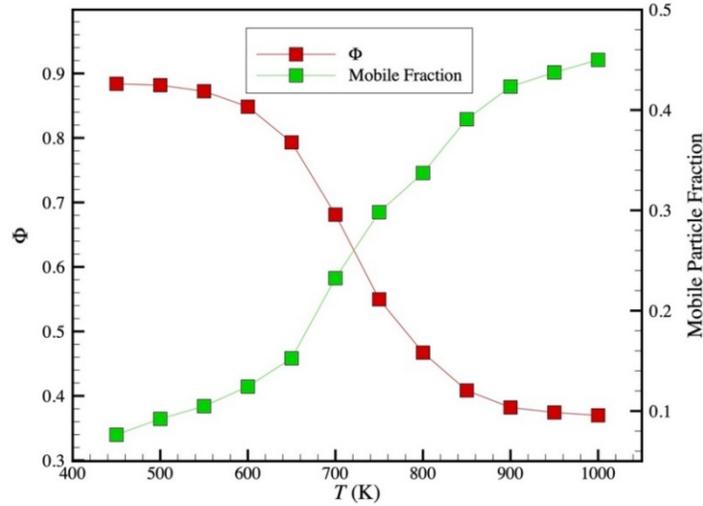

**Figure S1.** Fraction of mobile particles defined as the ratio of the number of mobile atoms to the total number of the atoms in the system, where the mobile atoms are the ones with displacement between (1.6 and 4.0) Å within the mobile particle lifetime, compared to the order parameter for the extent of string-like collective exchange motion, $\Phi$, which is defined as the ratio of number of atoms involved in the string-like motion to the number of mobile atoms. We note that the inflection point of these properties occurs close to the lambda transition temperature of this material $T_\lambda \approx 750$ K, where the response functions such as the specific heat exhibit extrema corresponding to some sort of order-disorder transition. We discuss the determination of $T_\lambda$ and the other characteristic temperatures of glass-formation of our Al-Sm metallic glass-forming liquid in the SI in our companion paper. [3] Interestingly, we found that the string length $L$ and $\Phi$ (unpublished data) exhibited their string polymerization transitions, defined by an inflection point in these properties close to the corresponding lambda transition in our previous simulations of superionic of $UO_2$ [17] The excess entropy of both our Al-Sm glass-forming liquid and superionic $UO_2$ exhibit an inflection point in the vicinity of $T_\lambda$. All these observations suggest that the λ-transition in these materials has its origin in the string polymerization transition, a well-defined rounded thermodynamic transition.

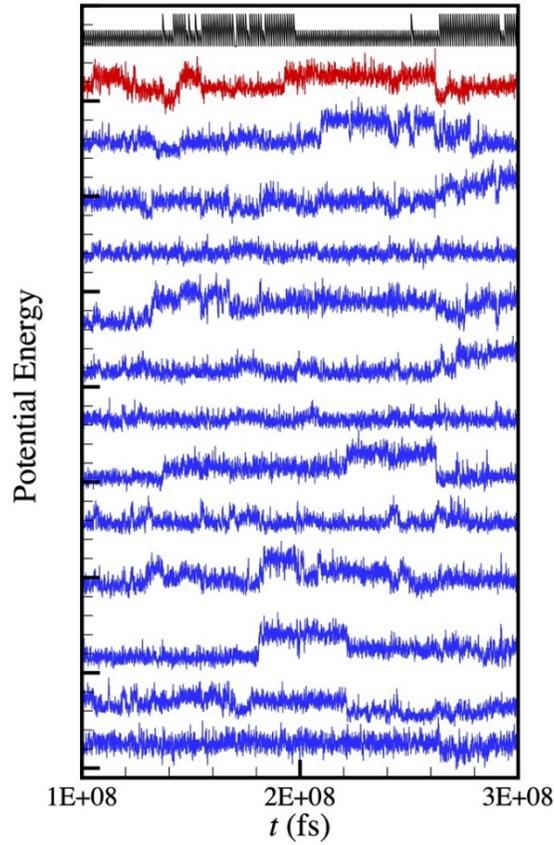

**Figure S2.** Potential energy fluctuations about a breather particle involved in a cooperative exchange event over the simulation timescale shown. The times series at the top show the coordination jumps (black) and potential energy jumps (red) of the breather. The blue curves show the potential energy fluctuations of atoms in the first coordination shell.

We next consider how the shear modulus $G$ in the glass state below $T_\lambda$ changes with the concentration of the mobile particles since the variation is expected to be large in the Granato model framework [4-6] in Fig. S3. Instead of the expected exponential dependence [4-6] of $G$ on the concentration of the excitations, we see a linear decrease, where $G$ appears to vanish near $T_\lambda$ (see Fig. S1). On the other hand, the concentration dependence is indeed quite large, in qualitative accord with the Granato theory. We note also that $G$ for crystalline Ni was observed to have a linear dependence on the interstitial concentration. [16] A linear variation of $G$ with defect concentration at lower concentrations is expected from lattice rigidity percolation theory [18-20] (See Eqn. 8 of Ref. [20]) so this may be a general result.

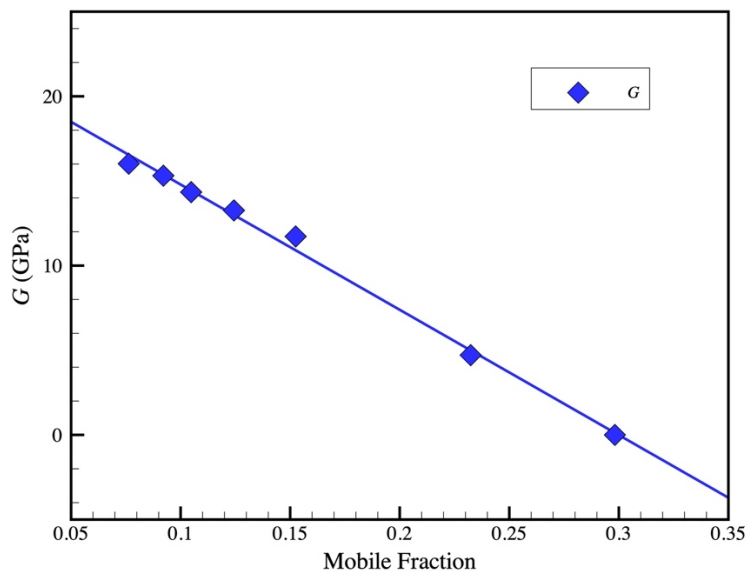

**Figure S3.** Shear modulus as function of the mobile particle concentration. The blue line is the linear fitting between the mobile fraction and the shear modulus $G$. The predicted slope for a low interstitial concentration in Granato theory is 23 so we would anticipate a slope of about unity based on our approximate relationship between the mobile and breather particle concentrations. The observed slope is somewhat larger value of about 7.9, but this is of the general expected order of magnitude from the Granato theory.

Finally, we examine how the concentration of the mobile particles influence the amplitude of the Johari-Goldstein (JG) relaxation process, defined as the height of the peak in measurements of the frequency dependent loss Young's modulus, $E''(\omega)$. While the Granato model [4-6] has nothing directly to say about the JG-relaxation process, we found in paper I [3] that the lifetime of the mobile particle clusters was directly related to the JG relation time so we might expect the amplitude of the JG relaxation process to be related to the number of mobile particles in this system. We indeed find evidence supporting this supposition in Fig. S4.

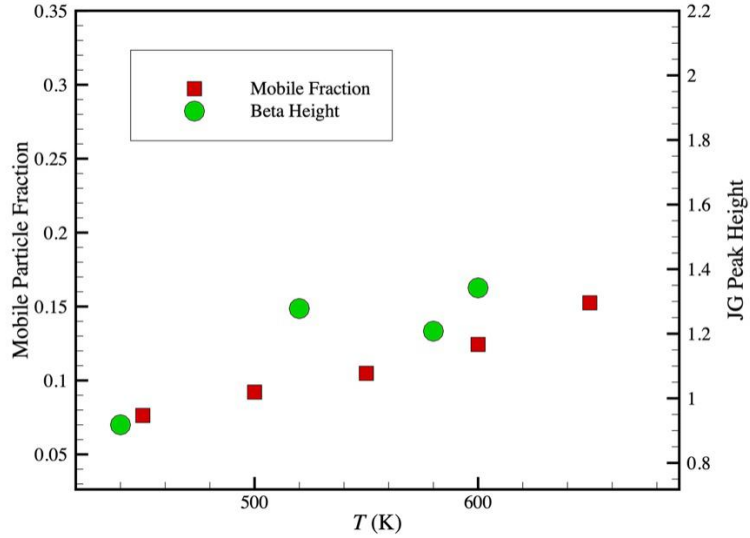

**Figure S4.** Variation of the Johari-Goldstein relaxation peak height [21] and the fraction of mobile particles. The peak height directly reflects changes in the mobile particle and presumably the breather particle concentration.

## C. Elastic Modulus Calculation

We performed uniaixal tensile tests at different temperatures of our Al-Sm metallic glasses to estimate some elastic properties, such as Young's modulus and Poisson's ratio. In particular, a constant true strain rate of $1 \times 10^7$ /s was applied in y-direction, and zero stresses were applied in x- and z-directions during the deformation, and periodic boundary conditions were maintained in all directions. The stress strain curves at different temperatures were plotted in Figure S5, where the slopes in the linear region yielded Young's modulus and the ratio of strain perpendicular to the loading direction to the strain along loading direction allowed us to calculate the Poisson's ratio. We also calculated the isothermal compressibility κ as function of $T$, defined as,

$\kappa = (<V^2> - <V>^2)/(k_B T <V>)$, where $V$ is the volume of the system and $<...>$ denotes NPT ensemble average. The bulk modulus $K$ is the defined by the reciprical of the isothermal compressibility.

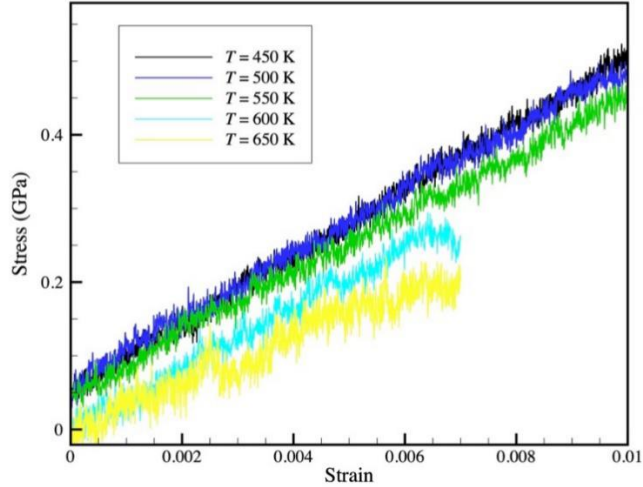

**Figure S5.** Stress strain curves of uniaxial tensile tests at different temperatures in our Al-Sm metallic glasses. The slope in the linear region (usually less than 0.5 %) yields Young's modulus and the relative simulation cell changes in loading direction and perpendicular directions allow us to calculate the Poisson's ratio.

## D. Determination of Fast Relaxation Times

We also examined the "collision time" or "Einstein time", $\tau_E$, another fast relaxation time related to the average collision rate. In particular, we fitted the initial decay of the velocity autocorrelation function to $VACF(t) = <v(0)^2> (1 + (t/\tau_E)^2/2 + O(t^3))$, as shown the solid red line in Figure S6(a). The "Einstein time" $\tau_E$ as a function of $T$ is shown in the inset of Figure S6(a). The reciprocal of $\tau_E$ is a measure of the average collision rate, which has a typical magnitude on the order of magnitude of $10^{-14}$ s in our metallic glass, a time on the order of $\tau_f$ / 10.

One of the other convention ways to define the decay of fast beta relaxation time is to define the cage size and then determine the decorrelation time. Figure S6(b) shows a typical mean square atomic displacement $<r^2>$ as a function of simulation time at $T$ = 900 K in our Al-Sm metallic glass. The mean square atomic displacement $<r^2>$ exhibits a well-defined plateau after a particular decorrelation time characterizing the crossover from ballistic to caged atom motion and

$<r^2>$ at long times is diffusive. Since the logarithmic derivative $d(\ln\langle r^2(t)\rangle)/d(\ln t)$ exhibits a minimum on the time scale of particle caging, $\tau_{fb}$, (see the inset of Figure S6(b)) we can then define the cage size by $r_{cage} \equiv \langle r^2(\tau_{fb})\rangle^{1/2}$. In Figure S7, we compare the cage size determined by the minimum of $d(\ln\langle r^2(t)\rangle)/d(\ln t)$ and by the mean first passage time [22]. Apparently, the cage size estimated by the mean first passage time is larger than the estimation by $<u^2>$ at low temperature, however, they are converging near the onset temperature, $T_A$

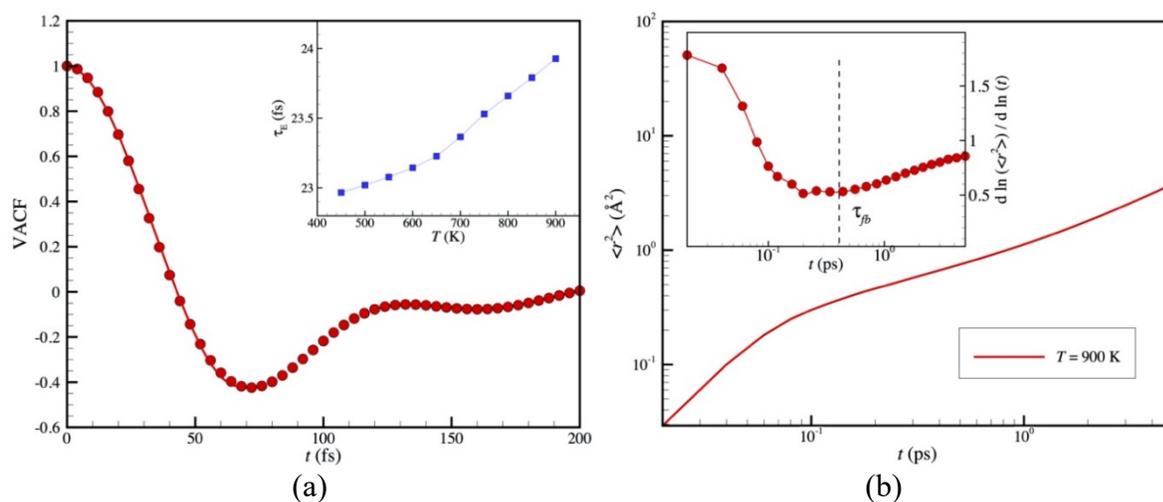

(a)                              (b)

**Figure S6.** (a) Initial decay of velocity autocorrelation function to obtain the collision time. The inset shows the mean "collision time" as a function of temperature. (b) A typical mean square displacement as a function time at $T$ = 900 K. The inset shows that the logarithmic derivative of $<r^2>$ exhibits a minimum on a time scale comparable to $\tau_{fb}$, a time on the order of a ps.

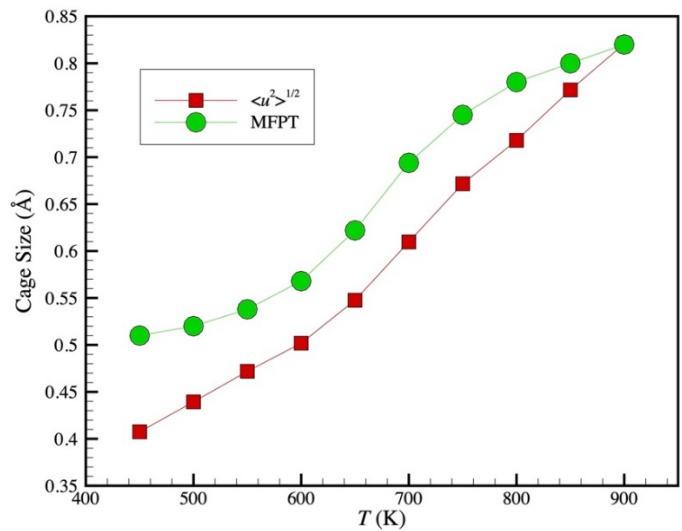

**Figure S7.** Comparison of the mean cage size determined by mean first passage time analysis[18] and $<u^2>^{1/2}$.

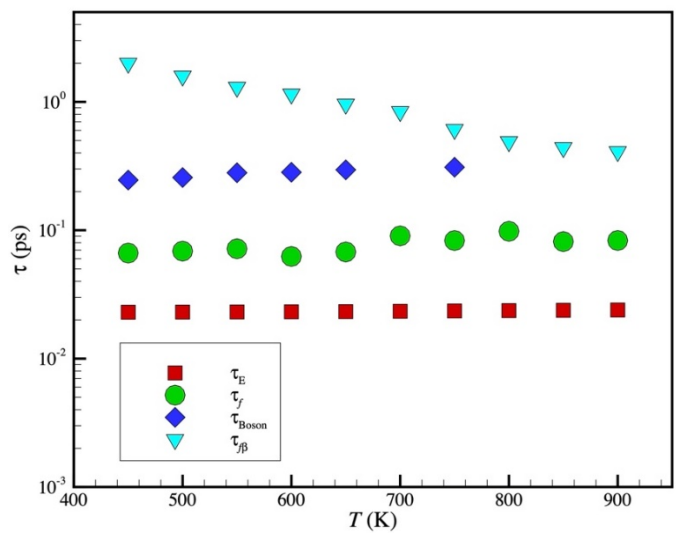

**Figure S8.** Summary of fast relaxation times as a function of temperature. $\tau_E$ is the "Einstein time", which is obtained by fitting the initial decay of the velocity autocorrelation function, as shown in Figure S6(a). $\tau_{Boson}$ is defined as $\tau_{Boson} \equiv 1/(2\pi\,\omega_B)$, where $\omega_B$ is the Boson peak frequency. $\tau_{f\beta}$ is the cage decorrelation time, which is the time that logarithmic derivative $d(\ln\langle r^2(t)\rangle)/d(\ln t)$ exhibits a minimum, as shown in Figure S6(b). $\tau_f$ is the fast beta relaxation time, obtained by fitting the self-intermediate scattering function. All of the fast relaxation times show weak temperature dependence in the current glass-forming liquid.

### E. Attempt to Fit the Structural Relaxation Time to the VFT Relation

The Vogel–Fulcher–Tammann (VFT) has often been used to define $T_g$ and $T_o$, however as noted in the SI in our companion paper [3], the applicability of this expression is highly uncertain for fluids undergoing a fragile to strong transition. If we fit our relaxation time data for the Sm-Al metallic glass to the VFT equation[23], $\tau_\alpha / \tau_0 \sim \exp[D_f T_0 / (T - T_0)]$ we may be estimated the $T$ at which $\tau_\alpha$ formally diverges $T_o$ and $T_g$ may be defined as the $T$ at which $\tau_\alpha$ extrapolates to 100 s. It is evident from Fig. S9 that the VFT equation provides a poor description of our relation data for $T$ greater than our estimates of $T_c$ above, and even if we attempt fitting our $\tau_\alpha$ data to the VFT in a higher $T$ regime, we estimate $T_g$ to equal $T_{g,VFT}$ = 603 K and $T_{o,VFT}$ = 565 K. This characteristic temperature estimate is evidently questionable for systems undergoing a fragile to strong transition.

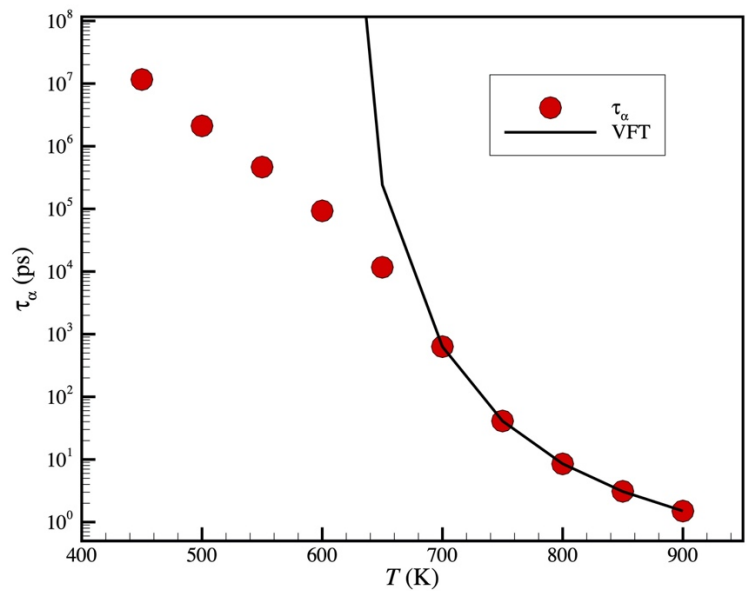

**Figure S9.** VFT fitting of structural relaxation time $\tau_\alpha$.

## F.    Signature of the Boson Peak in the Non-Gaussian Parameter

Previous simulations by Binder and coworkers [24] on silica, another fluid exhibiting a fragile to strong transition and an appreciable Boson peak, indicated that the non-Gaussian parameter $\alpha_2$ exhibited a kink at times on the order of the timescale of the Boson peak, $\tau_{Boson}$ (see Fig. S.8). This struck as a fascinating suggestion and we correspondingly checked to see if such a feature arises in our metallic glass and, if so, whether it might correspond to the Boson peak time. In Fig. S10, we see there is indeed a feature in $\alpha_2$ which indeed closely coincides with the Boson peak time. The concerted motion underlying the Boson peak evidently makes a small signature in the particle displacement dynamics.

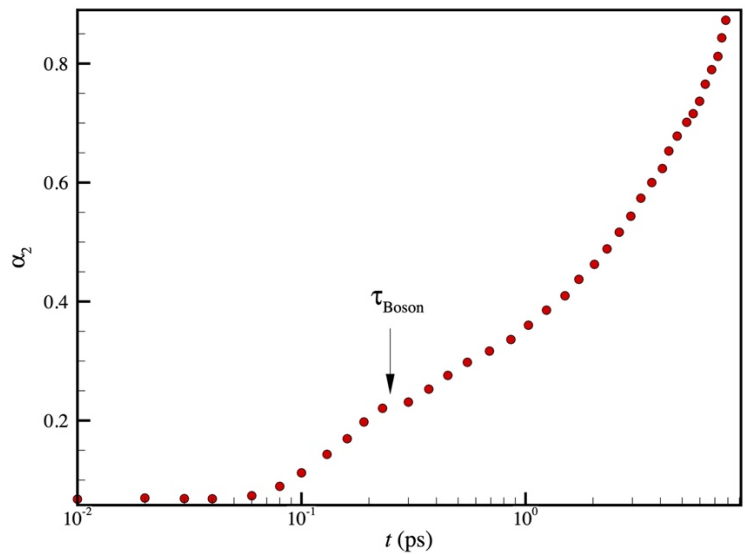

**Figure S10.** Non-Gaussian parameter at short time in an Al90Sm10 alloy at $T$ = 500 K, where a kink apparently appears at the position coinciding with Boson peak time at the same temperature.

## G. Movie of Cooperative Movement of Stringlet

The animation shows typical cooperative motion of a stringlet involving four atoms in a time period of 100 ps at $T = 450$ K. The dashed line circle represents the "cage size" of each atom at the initial time. It is evident that the atoms repeatedly jump back-and-forth to the original 'cage', making a reversible collective movement.